\documentclass[11pt]{article}
\usepackage[latin1]{inputenc}
\usepackage{amsmath}
\usepackage{amsfonts}
\usepackage{amssymb}
\usepackage{graphicx}
\usepackage{footnote}
\usepackage{float}
\usepackage[titletoc,title]{appendix}
\usepackage{subfigure}

\DeclareGraphicsExtensions{.bmp,.png,.pdf,.jpg,.eps}

\usepackage[colorlinks]{hyperref}
\hypersetup{
    colorlinks=true,
    linkcolor=blue,
    filecolor=magenta,  
    urlcolor=[rgb]{0.00,0.00,0.50},    
    citecolor=red,
    }

\usepackage{url}

\advance \textheight by \topskip
\usepackage{amsmath}
\usepackage[english]{babel}
\makeatletter
 \@addtoreset{equation}{section}
\makeatother

\usepackage{amsmath}
\numberwithin{equation}{section}

\advance \textheight by \topskip
\usepackage{amsmath}
\usepackage[english]{babel}

\def\be{\begin{equation}}
\def\ee{\end{equation}}

\def\A{\mathbb A}

\def\Z{\mathbb Z}
\def\R{\mathbb R}

\def\N{\mathbb N}

\usepackage{todonotes}

\begin{document}

\title{
{\bf 
Rational deformations of
conformal mechanics 
 }}

\author{{\bf Jos\'e F. Cari\~nena${}^a$, Luis Inzunza${}^b$ and Mikhail S. Plyushchay${}^b$} 
 \\
[8pt]
{\small \textit{
${}^a$Departamento de F\'{\i}sica Te\'orica, 
Universidad de Zaragoza, 50009 Zaragoza, Spain}}\\
{\small \textit{ ${}^b$Departamento de F\'{\i}sica,
Universidad de Santiago de Chile, Casilla 307, Santiago 2,
Chile  }}\\
[4pt]
 \sl{\small{E-mails:  \textcolor{blue}{jfc@unizar.es}, 
 \textcolor{blue}{luis.inzunza@usach.cl},
\textcolor{blue}{mikhail.plyushchay@usach.cl}
}}
}
\date{}
\maketitle

\begin{abstract}
We study deformations  of the  quantum conformal mechanics of  De Alfaro-Fubini-Furlan
with rational additional potential term  generated 
 by applying the generalized Darboux-Crum-Krein-Adler  transformations to the 
quantum harmonic oscillator  and by using the method of dual schemes and mirror 
diagrams.
In this way we obtain  infinite families 
of  isospectral and non-isospectral  deformations of the conformal mechanics model 
with special values of the coupling constant $g=m(m+1)$, $m\in\N$, in the inverse square potential term, 
and
for each  completely isospectral or gapped deformation 
given by a mirror diagram, we identify  the  sets of  the spectrum-generating ladder operators 
which encode and coherently reflect  its fine spectral  structure. 
Each pair of these operators generates a nonlinear deformation of the conformal 
symmetry, and their complete sets 
pave  the way for investigation of the associated 
superconformal symmetry deformations.

\end{abstract}

\vskip.5cm\noindent

\section{Introduction}

Conformal mechanics of De Alfaro, Fubini and Furlan was introduced
and studied  as a (0+1)-dimensional conformal field theory \cite{deAFF}.
This system corresponds to a simplest two-body  case of the 
$N$-body  Calogero model 
\cite{Calogero1,Calogero2,Perel} with eliminated center of mass 
degree of freedom. 
The same quantum  harmonic oscillator model with a centrifugal  barrier \cite{deAFF}
appeared in the literature  later under the name of the  isotonic 
oscillator \cite{WeiJor}.
It found interesting applications, particularly, 
in the context of the physics of anyons \cite{LeiMyr,MacWil,Polych}. 
For some references devoted to investigation 
of different aspects of this model and its generalizations see 
\cite{AkuPas,FubRab,IvaKriLev1,IvaKriLev2,FreMen,Wyl,ACMP,CaPeRa}.
Some time ago a new wave of interest to this model and 
its supersymmetric generalizations appeared in the context 
of the AdS/CFT correspondence \cite{AdS/CFT1,AdS/CFT2,AdS/CFT3}
after observation that the dynamics 
of a particle near the horizon of the extreme Reissner-Nordstr\"om 
black hole is governed by some relativistic  type of 
conformal mechanics 
\cite{CDKKTV,AIPT,GibTow,MicStr,BPMSV,Papado,PioWal,FedIvaLec,Gonera}.
Recently, the model also attracted attention in connection
with the confinement problem in QCD \cite{deTerDosBro,BrTeDoLo}.
The isotonic oscillator also caused
a notable  interest in investigations 
related to  
exceptional orthogonal polynomials 
\cite{Dub,CPRS,FelSmi,GUKM1,Ques1,
Grandati1,Mar1,CarPly1}.

 The method of the Darboux-Crum-Krein-Adler (DCKA)
transformations  \cite{Darb,Crum,MatSal,Krein,Adler}
provides a simple mechanism for construction
of the rational extensions 
of  the quantum harmonic 
oscillator (described by the potential which is composed from 
the quadratic term and additionally generated rational term) with
any preassigned almost equidistant gapped spectrum 
containing  arbitrary finite even number of missing energy levels in it
\cite{Adler,Obl}.
The same method  allowed recently 
to identify the set of the ladder operators for
rationally extended quantum harmonic oscillator systems
that  detect and coherently reflect all their spectral peculiarities
and form the sets of the spectrum-generating 
operators \cite{CarPly2}. 

\begin{itemize}
\item
 In the present  paper, we address the problem of construction 
of the complete sets of the spectrum-generating ladder operators for
rational deformations of the conformal mechanics
characterized by  special values of the coupling constant $g=m(m+1)$, $m\in\N$,
in the inverse square potential term.
Each pair of such ladder operators provides us
with a certain nonlinear deformation 
of the conformal symmetry, and their complete sets 
pave  the way for investigation of the associated 
superconformal symmetry deformations.
\end{itemize}

The conformal mechanics model with the indicated  values 
of the coupling constant in the inverse square potential term
are especially interesting from the point of view of  applications to the physics of anyons
where the cases of $g=m(m+1)$ describe the 
 relative dynamics of the bosons 
\cite{LeiMyr,MacWil,Polych,MatPly}.
Another reason is that the potentials $m(m+1)/x^2$ without the confining 
harmonic term are related  to the hierarchy of the Korteweg-de Vries (KdV) equation
within a framework  of a broader picture 
in which the Calogero systems govern the dynamics of the moving
poles of the rational solutions to the KdV equation 
\cite{AirMcKMos,AdlMos,GorNek}.
\vskip0.1cm

We start  from a certain set of physical and non-physical eigenstates of the quantum 
harmonic oscillator
selected as the seed states for the DCKA transformations, 
and  obtain infinite families  of rational  extensions 
 of the isotonic oscillators for both isospectral and non-isospectral cases
 with arbitrary (odd or even) number of missing energy levels in the gaps.  
For this, we develop and apply the method
of the mirror diagrams allowing us to identify  
complementary  sets of the seed states which
produce  the same system modulo a relative displacement 
of the spectrum. Having  two dual schemes for the same system,  
we
construct
the sets  of the ladder operators 
which 
generate nonlinearly deformed  conformal symmetry and 
coherently reflect all the spectral properties of  each 
isospectral and non-isospectral deformation.

\vskip0.1cm

The paper is organized as follows. 
The next section  is 
devoted to the construction and analysis 
of the quantum conformal mechanics  systems
by applying  
the Darboux-Crum (DC) transformations 
to the half-harmonic oscillator. 
One of the 
two dual 
families 
of transformations  we use is based on 
selection 
of physical eigenstates of the quantum harmonic 
oscillator as the seed states. Another family of Darboux transformations 
 employs  non-physical eigenstates generated 
from physical eigenfunctions  by a spatial Wick rotation.
 The same two types of the Darboux transformations 
 are employed  for the construction of the ladder operators 
 of the isotonic oscillator systems.
We also observe there how  ladder operators can be  generated 
from the two supersymmetric schemes corresponding 
to the broken and unbroken phases. 
In Section \ref{SectionDual}  we 
describe the construction of the dual schemes of  
the DCKA
transformations by generalizing the relations 
from  Section \ref{SectionIsotonic}.
We show there that the dual schemes can conveniently be presented
by a mirror diagram, in which they are related 
by a kind of a ``charge conjugation". 
The dual schemes and mirror diagrams 
lie in the basis of constructions of  the next two sections.
In Section \ref{SectionIso}  we
construct  infinite families of 
the completely  isospectral rational deformations
of the conformal mechanics  systems 
and describe their ladder operators.
In Section \ref{SecGapped} we describe 
the construction of  rational deformations 
with  any preassigned gapped
spectrum organized in an arbitrary number of 
the lower-lying ``valence" bands
of arbitrary size and  infinite equidistant  band.
We also consider there the construction 
of the 
pairs of ladder operators which reflect different 
properties of the spectrum, and then identify  their combinations 
 that form the sets of the spectrum-generating 
 operators. Each such a pair of the ladder operators
 generates a certain nonlinear deformation of the conformal
 symmetry. In Section \ref{SectionSummary}
 we summarize and discuss 
 the results,
 and indicate some problems
 that could be interesting for further investigation.

\section{
Quantum conformal mechanics systems 
}\label{SectionIsotonic}

Consider the quantum harmonic oscillator system
described by the Hamiltonian  
defined  by the differential operator
$L^{\rm osc}=-\frac{d^2}{dx^2}+x^2$ with domain $L^2(\R, dx)$. 
The solutions to the spectral  problem 
$L^{\rm osc}\psi_n=E_n\psi_n$
are given by  the normalizable 
eigenfunctions 
$\psi_n(x)=H_n(x)e^{-x^2/2}$ corresponding to the energy
values  $E_n=2n+1$, $n=0,1,\ldots$, where $H_n(x)$ are the Hermite polynomials.
Here and in what follows we do not preoccupy ourselves with normalization
of the states
and specify wave functions modulo multiplication by a
nonzero complex number.
A simple change $x \rightarrow ix$, 
see \cite{Dub,FelSmi,
CarPly1,Obl,CMPR}, 
provides us with non-physical (non-normalizable) solutions
 $\psi_n^{-}(x)=\mathcal{H}_n(x)e^{x^2/2}$ that 
correspond to  eigenvalues $E_{n}^-=-E_n$,
where $\mathcal{H}_n(x)=(-i)^nH_n(ix)$.
These not physically acceptable  states 
 will play an important role 
 in the construction of the rational  
deformations
of the quantum mechanics  systems.

The confining AFF model \cite{deAFF} is 
characterized  by the $\mathfrak{sl}(2,\R)$  conformal symmetry 
\begin{equation}
\label{sl2R}
[L^{\rm{iso}}_g, \mathcal{C}^\pm_g]=\pm4\mathcal{C}^\pm_g\,,
\qquad [\mathcal{C}^-_g,\mathcal{C}^+_g]=8L^{\rm{iso}}_g\,, 
\end{equation}
where $L^{\rm{iso}}_g$ is the Hamiltonian given by
a  Schr\"odinger type differential operator
\begin{equation} \label{defisog}
L^{\rm{iso}}_g=-\frac{d^2}{dx^2}+x^2+\frac{g}{x^2}\,, \qquad
g\in\R\,,
\end{equation}
defined in a domain  
$\{\phi\in L^2((0,\infty),dx)\mid \phi(0^+)=0\}$,
and $\mathcal{C}^\pm_g$ are the  spectrum-generating ladder operators
which we specify below. 
When $g\geq -1/4$, one can replace 
$g=\nu(\nu+1)$ with 
 $\nu\in\R$. For $g>-\frac{1}{4}$,
  the factorization can be done in two ways 
 because $g(\nu)=g(-(1+\nu))$.
We  are interested here  in the case 
when $\nu$ takes integer values,
and  write
\begin{equation} \label{defiso}
L^{\rm{iso}}_m=-\frac{d^2}{dx^2}+x^2+\frac{m(m+1)}{x^2}\,,
\qquad m=0,1,\ldots.
\end{equation}
The first member $L^{\rm iso}_0$ of this infinite family corresponds to 
a \emph{half-harmonic oscillator} (on the 
half-line $x> 0$)
with a non-penetrable infinite barrier put at $x=0$, 
i.e. to  the quantum system given by the potential 
$V(x)=+\infty$ for $x\leq 0$ and $V(x)=x^2$ for $x>0$.
Its physical eigenstates $\psi_{0,l}$ and eigenvalues $E_{0,l}=4l+3$, $l=0,1,\ldots$,
correspond to those of the quantum harmonic 
oscillator with odd index $n=2l+1$\,:
$\psi_{0,l}(x)=\psi_{2l+1}(x)$, 
for $x\in(0,\infty)$,
$E_{0,l}=E_{2l+1}$.
Hereafter with an abuse of notation we write that two functions with a different 
domain are
equal when both coincide on the common domain, 
and in the preceding case we simply write 
$\psi_{0,l}=\psi_{2l+1}$.
\vskip0.1cm

The eigenstates and eigenvalues of the isotonic (radial)
oscillator $L^{\rm iso}_m$ with $m\geq1$ can be found, particularly,  
by applying the DC  transformations to
the half-harmonic oscillator system $L^{\rm iso}_0\equiv L_0$. 
Taking the first $m$  eigenfunctions 
$\psi_{0,0},\ldots,\psi_{0,m-1}$,
$m=1,2,\ldots$, 
of $L_0$ 
as the seed states for the DC  transformation,  see  Appendix \ref{subSecDC},
we generate the quantum Hamiltonian
\begin{equation}\label{LmWron}
L_m\equiv  -\frac{d^2}{dx^2}+x^2-2 
\big(\ln W(1,3,\ldots,2m-1)\big)''=
L^{\rm iso}_m+2m\,.
\end{equation} 
Here and in what follows we  use  a  compact notation 
$n=\psi_n(x)$, 
$-n=\psi^-_n(x)$, 
$\widetilde{n}=\widetilde{\psi_n(x)}$, 
$\widetilde{-n\,}=\widetilde{\psi^-_n(x)}$
for the corresponding physical and non-physical eigenstates
of the quantum harmonic oscillator, 
where 
$\widetilde{\psi_n(x)}$
is an eigenfunction  linearly independent 
from a solution $\psi_n(x)$ of the 
corresponding stationary Schr\"odinger equation,
see Eq. (\ref{Atildepsi}).
According to this notation, 
$W(1,3,\ldots,2m-1)$ is the Wronskian 
of the odd eigenfunctions 
$\psi_1=\psi_{0,0}$, 
$\psi_3=\psi_{0,1},\ldots,\psi_{2m-1}=\psi_{0,m-1}$ of the quantum 
harmonic oscillator.
Hereafter 
 we identify 
an eigenfunction of a 
 differential operators $L$ 
as  a physical eigenstate of the corresponding 
Hamiltonian $H$ when it
satisfies the integrability and boundary 
conditions.
Accordingly, 
a non-physical eigenstate 
is not a
true  eigenstate of the 
Hamiltonian operator.

Given the physical character 
of the eigenstates of $L_0=L^{\rm iso}_0$ that
 we used as seed states 
for the DC transformation  (\ref{LmWron}), 
the corresponding energy levels are absent in the 
spectrum of the generated system  $L_m$.
The neighbour members $L_{m-1}$ and $L_m$ of the family of the 
differential operators and their corresponding 
Hamiltonians
are factorized and intertwined by the first order operators
\begin{equation}
\label{4}
A_m^{-}=\frac{d}{dx}+x-\frac{m}{x}\,, \qquad
A_{m}^{+}=-\frac{d}{dx}+x-\frac{m}{x}\,,
\end{equation}
in terms of which we have
\begin{eqnarray}
&A_{m}^{+}A_{m}^{-}=L_{m-1}-E_{2m-1}\,,\qquad
A_{m}^{-}A_{m}^{+}=L_{m}-E_{2m-1}\,, &\\
&A_{m}^{-}L_{m-1}=L_{m}A_{m}^{-}\,,  \qquad
A_{m}^{+}L_{m}=L_{m-1}A_{m}^{+}\,. &
\end{eqnarray}
Here  
$E_{2m-1}=4m-1=E_{m-1,0}$ corresponds 
to the eigenvalue of the eigenfunction
$\psi_{m-1,0}=x^{m}e^{-x^2/2}$
that describes the ground state of the system $L_{m-1}$.
According to Eq. (\ref{defApsi}), the state  $\psi_*=\psi_{m-1,0}$
 generates the operators (\ref{4}).
Any eigenstate $\psi_{m,E}$ of energy $E\neq E_{m-1,0}$ of the system $L_m$ can be generated 
from the state $\psi_{m-1,E}$ that is an eigenstate of $L_{m-1}$ of the same energy,
$\psi_{m,E}=A^-_m\psi_{m-1,E}$. The 
non-physical eigenstate $(\psi_{m-1,0})^{-1}$ of $L_m$ of energy
$E=E_{m-1,0}$ is generated 
 by applying the operator $A^-_m$ to  the 
 non-physical eigenstate 
 $\widetilde{\psi_{m-1,0}}$ of $L_{m-1}$.
 
 The factorization and intertwining relations satisfied by operators (\ref{4}) 
 can be translated into the language 
 of supersymmetric quantum mechanics by  taking 
the $2\times 2$ diagonal matrix
  \be\label{mathcalLm}
 \mathcal{H}_{m}^e=\text{diag}\,( \mathcal{H}^{e,+}_{m}\equiv L_{m}-E_{2m-1}
\,,\,
 \mathcal{H}^{e,-}_{m}\equiv L_{m-1}-E_{2m-1})\,,\qquad m=1,\ldots,
 \ee
 as the Hamiltonian operator.  The antidiagonal supercharge operators are defined by
 \be\label{mathcalQma}
 \mathcal{Q}_{m}^{e,1}=
\left(
\begin{array}{cc}
  0&    A^-_m  \\
 A^+_m &   0     
\end{array}
\right),
\qquad \mathcal{Q}_{m}^{e,2}=
i\sigma_3\mathcal{Q}_{m}^{e,1}=
\left(
\begin{array}{cc}
  0&    iA^-_m  \\
 -iA^+_m &   0     
\end{array}
\right).
\ee
They generate the Lie superalgebra
of $\mathcal{N}=2$ supersymmetry,
\be
[\mathcal{H}^e_m,\mathcal{Q}_{m}^{e,\alpha}]=0\,,\qquad
\{\mathcal{Q}_{m}^{e,\alpha},\mathcal{Q}_{m}^{e,\beta}\}=2\delta^{\alpha,\beta}
\mathcal{H}^e_m\,,\quad
\alpha,\beta=1,2\,.
\ee
The matrix Hamiltonian operator $\mathcal{H}_m^e$  
has the singlet 
ground state 
$\Psi_{0,m}=(0,\psi_{m-1,0})^{t}$  of zero energy
that  is annihilated by both supercharges
$\mathcal{Q}_{m}^{e,\alpha}$, and  so,
the extended system (\ref{mathcalLm}), (\ref{mathcalQma}) 
corresponds to the case of the exact, 
 unbroken supersymmetry.

 In correspondence with the described picture, 
 physical eigenstates $\psi_{m,l}$, 
 of $L_m$ of energies $E_{m,l}=4(m+l)+3$, 
 $l=0,1,\ldots$,  can be generated from the odd eigenstates 
 $\psi_{2(m+l)+1}$  of the quantum harmonic oscillator
 of the same energies $E_{2(m+l)+1}=4(m+l)+3$, 
 \begin{equation}\label{AAm}
\psi_{m,l}=\mathbb{A}_m^{-}\psi_{2(m+l)+1}\,,\qquad
\mathbb{A}_m^{-}=A_m^{-}A_{m-1}^{-}\ldots A_1^{-}\,.
\end{equation}
In the particular case of $l=0$, relation (\ref{AAm})
gives  the ground state of the system $L_m$, 
$\psi_{m,0}=\mathbb{A}_m^{-}\psi_{2m+1}=C x^{m+1}e^{-x^2/2}$.
The operator $\A_m^-$ annihilates all the $m$ lowest odd  eigeinstates $\psi_{2n-1}$
of 
the quantum harmonic oscillator  with $n=1,\ldots, m$.
If $\psi_E$ is any physical or non-physical eigenstate of the quantum harmonic 
oscillator with $E\neq E_{2n-1}=4n-1$, $n=1,\ldots, m$,
Eq.  (\ref{AAm}) can be generalized as follows\,:
\begin{equation}\label{psiEpsi+}
\psi_{m,E}=\A_m^-\psi_{E}\,,\qquad
L_m\psi_{m,E}=E\psi_{m,E}\,.
\end{equation}
 Here $\psi_{m,E}$
 is an eigenstate of $L_m$
 of the same, physical or non-physical, nature
 excepting the case when $E=E_{2l}$, $l=0,1,\ldots$, 
 and $\psi_E=\psi_{2l}$. 
 The physical eigenstate $\psi_{2l}$ of the quantum harmonic 
 oscillator being a non-vanishing at $x=0$ even function
  is mapped into non-physical eigenstate 
 of $L_m$ that is singular at $x=0$. 
 Notice that the relation (\ref{psiEpsi+}) can be presented equivalently
in a Wronskian form,
$\psi_{m,E}=
{W(1,3,\ldots,2m-1,\psi_{E})}/{W(1,3,\ldots, 2m-1)}$.
Using the factorization and intertwining relations satisfied 
by operators (\ref{4}), 
we find the  
relations 
\begin{equation}\label{AmAmgen}
\mathbb{A}_m^{+}\mathbb{A}_m^{-}=\prod_{j=1}^{m}(L_0-E_{2j-1})\,,\qquad
\mathbb{A}_m^{-}\mathbb{A}_m^{+}=\prod_{j=1}^{m}(L_m-E_{2j-1})\,,
\end{equation}
and 
$\A_m^-L_0=L_m\A_m^-$, 
$\A^+_mL_m=L_0\A^+_m$ 
for the higher order 
differential operators $\A_m^-$, 
$\A_m^+=(\A_m^-)^\dagger$. 
Eq. (\ref{AmAmgen}) 
is useful, particularly,  
to establish the exact relationship
between  eigenfunctions (\ref{AAm}) 
of the isotonic oscillator given in terms of the Hermite polynomials
and their standard representation 
in terms of the generalized Laguerre polynomials,
see Appendix \ref{AppHerLag}.

The method of dual schemes and mirror diagrams
 described below in Section
\ref{SectionDual}, allows us to find alternative ways 
to generate the isotonic oscillator system $L_m$
by means of the DC transformations.  
This can be done, particularly, by using the
odd non-physical  eigenstates $\psi^{-}_{2l+1}$ of the quantum harmonic oscillator 
instead of its physical odd eigenstates $\psi_{2l+1}$.
The  states $\psi^{-}_{2l+1}$  are also non-physical  eigenstates  of 
the half-harmonic oscillator $L_0=L^{\rm iso}_0$
of eigenvalues  $-4l-3$, and  instead of (\ref{LmWron}) we obtain 
\begin{equation}\label{LmWron-4m}
L_m-4m=-\frac{d^2}{dx^2}+x^2-2\frac{d^2}{dx^2}\big(
\ln W(-1,-3,\ldots,-(2m-1))\big)=L^{\rm iso}_m-2m\,.
\end{equation}
We have here the alternative representation 
for eigenfunctions (\ref{psiEpsi+}) of $L_m$\,:
$\psi_{m,E}=
{W(-1,-3,\ldots, -(2m-1),\psi_{E})}/{W(-1,-3,\ldots, -(2m-1))}$. 
Due to a non-physical nature of the seed states, in this case
 there are no removed energy values  in the spectrum of (\ref{LmWron-4m})
in comparison with the spectrum of the half-harmonic  oscillator
$L_0=L^{\rm iso}_0$.  The DC transformation of non-physical 
eigenstate $ \widetilde {\psi_{0, l}^{-}}=\widetilde{\psi^{-}_{2l+1}}$ of $L_0$, $l=1,\ldots,m$,
 produces  
a  non-physical eigenstate of 
the Hamiltonian operator $L_m-4$. As a result,  the system described
by the differential  operator 
$L_m-4$ turns out to be completely  isospectral to  $ L_0$. 
We shall show below that  this picture of DC transformations 
corresponds to 
the case of spontaneously broken supersymmetry.
The relative nonzero shift $-4m$ in (\ref{LmWron-4m}) in comparison with  
(\ref{LmWron}) is important and 
lies in the base of  the construction of the
ladder operators.

The shift $-4m$ in (\ref{LmWron-4m}) 
is  associated with the difference 
in the exponential factors in the structure 
of the physical states $\psi_n(x)$
of the quantum harmonic oscillator 
and its non-physical eigenstates $\psi^-_n(x)=C\psi_n(ix)$.
The structure of the Wronskian
(\ref{LmWron})  is
$W(1,3,\ldots,2m-1)=Cx^{\Delta_m}e^{-mx^2/2}$,
where 
 $\Delta_m={m(m+1)}/{2}$,
 see Appendix \ref{AppB},
while  the Wronskian in (\ref{LmWron-4m}) is obtained from
 it by the change $x\rightarrow ix$, 
 $W(-1,-3,\ldots,-(2m-1))=Cx^{\Delta_m }e^{mx^2/2}$.
 
 As in the case of the DC scheme based on physical 
 eigenstates of $L_0$, one can factorize and intertwine 
 the neighbour members in the family of 
 differential operators of
 the
 isotonic oscillator systems by the first order
 differential  operators 
 \begin{equation}\label{A-m}
A_{-m}^{-}=\frac{d}{dx}-x-\frac{m}{x}\,,\qquad
A_{-m}^{+}=-\frac{d}{dx}-x-\frac{m}{x}\,.
\end{equation}
These operators can be obtained
by means of  relation (\ref{defApsi})
with  $\psi_*=\psi_{m-1,0}^-$, where
$\psi_{m-1,0}^-=\psi_{m-1,0}(ix)=x^{m}e^{x^2/2}$
is
a  non-physical
eigenstate of $L_{m-1}$
of eigenvalue $-3$.
Note that 
the non-physical eigenstate $\psi_{m-1,0}^-$
of $L_{m-1}$ can 
be generated from the non-physical
eigenstate $\psi^{-}_{2m+1}$
of the quantum harmonic oscillator
if we take there $\psi_E=\psi^{-}_{2m+1}$.

If we formally put
 $m=0$ in $A_{m}^\pm$ and $A_{-m}^\pm$ , we obtain 
 the ladder operators  $(a^\pm)$ of the harmonic oscillator.
They are not of the same nature 
as  $A_{m}^\pm$ or $A_{-m}^\pm$  since
they intertwine harmonic oscillators with relatively displaced spectra,
and transform physical states of $L_0$ into its non-physical states\,:
$a^\pm \psi_{2n+1}=\psi_{2n+1\pm1}$. Nevertheless, they 
produce  the ladder operators $(a^\pm)^2$  of  $L_0$,
and will be used 
in the construction of mirror diagrams in the next section.

We have 
 the factorization relations
$A_{-m}^{-}A_{-m}^{+}=L_m^{\rm iso}+2m-1$
and
$A_{-m}^{-}A_{-m}^{+}=L_m-1$. 
Because of different constants of factorization,  
the operators 
$A_{-m}^{-}$ and $A_{-m}^{+}$, unlike $A_{m}^{-}$ and $A_{m}^{+}$,
intertwine the differential  operators $L_{m-1}$
and $L_m$ with an additional  nonzero shift,
\be
A_{-m}^{-}L_{m-1}=(L_m-\Delta E^{\rm iso})A_{-m}^{-}\,, \qquad 
A_{-m}^{+}L_{m}=(L_{m-1}+\Delta E^{\rm iso})A_{-m}^{+}\,.
\ee
Here $\Delta E^{\rm iso}=4$ is the difference 
between any two successive levels in the 
equidistant spectrum of 
$L_m$ for any value of $m$.  
As a consequence,  the application of  $ A_{-m}^{-}$ to a 
physical or non-physical  eigenstate $\psi_{m-1,E}$ of
$L_{m-1}$ of eigenvalue $E\neq -3$,  $L_{m-1}\psi_{m-1,E}=E\psi_{m-1,E}$,
produces a state $ A_{-m}^{-}\psi_{m-1,E}$ that is an eigenstate
of $L_m$ of eigenvalue $E+4$ having the same physical or non-physical nature.
Similarly, the application of  $ A_{-m}^{+}$ 
to an eigenstate $\psi_{m,E}$ of
$L_{m}$ of eigenvalue $E\neq +1$ 
produces the state $ A_{-m}^{+}\psi_{m,E}$  that is an eigenstate
of $L_{m-1}$ of eigenvalue $E-4$ having the
same,  physical or non-physical,
 nature as the state
$\psi_{m,E}$.
Besides, in accordance  with 
Eq. 
(\ref{Atildepsi}),
the application of $A_m^{-}$ and  $A_m^{+}$ 
 to the partner  states $\widetilde{\psi}$ of the 
 kernels of $A_m^{+}$  and $A_m^{-}$  reproduces 
 the corresponding non-physical  eigenstates of $L_{m}$ and $L_{m-1}$
 of energies $E=+1$ and $-3$, respectively.
 \vskip0.1cm

Analogously to  (\ref{mathcalLm}) and (\ref{mathcalQma}), 
we can translate the factorization and intertwining relations 
satisfied by the operators $A^-_{-m}$ and $A^+_{-m}$ into the language 
 of supersymmetric quantum mechanics by defining the extended 
 Hamiltonian and supercharge operators as follows\,:
 \be\label{mathcalLmbr}
 \mathcal{H}_{m}^b=\text{diag}\,( \mathcal{H}^{b,+}_{m}\equiv L_{m}-1
\,,\,
 \mathcal{H}^{b,-}_{m}\equiv L_{m-1}+3)\,,\qquad m=1,\ldots,
 \ee
 \be\label{mathcalQmarb}
 \mathcal{Q}_{m}^{b,1}=
\left(
\begin{array}{cc}
  0&    A^-_{-m}  \\
 A^+_{-m} &   0     
\end{array}
\right),
\qquad \mathcal{Q}_{m}^{b,2}=
i\sigma_3\mathcal{Q}_{m}^{b,1}=
\left(
\begin{array}{cc}
  0&    iA^-_{-m}  \\
 -iA^+_{-m} &   0     
\end{array}
\right).
\ee
They generate, again,  the Lie superalgebra
of  $\mathcal{N}=2$ supersymmetry,
\be
[\mathcal{H}_m^b,\mathcal{Q}_{m}^{b,\alpha}]=0\,,\qquad
\{\mathcal{Q}_{m}^{b,\alpha},\mathcal{Q}_{m}^{b,\beta}\}=2\delta^{\alpha,\beta}\mathcal{H}_m^b\,,
\quad \alpha,\beta=1,2\,.
\ee
But here the lowest energy eigenvalue 
of  the extended Hamiltonian 
operator $\mathcal{H}_m^b$  
is positive, $E^b_{0,m}=4m+2\geq 6$, 
and doubly  degenerate like its any higher energy level. 
The corresponding eigenstates 
are $\Psi_{0,m}^{+}=(\psi_{m,0},0)^t$ and 
$\Psi_{0,m}^-=(0,\psi_{m-1,0})^t$.
Any linear combination of the  states $\Psi_{0,m}^{+}$ and $\Psi_{0,m}^{-}$  is
not annihilated by the supercharges $\mathcal{Q}^{b,\alpha}_m$,
and the system  (\ref{mathcalLmbr}), (\ref{mathcalQmarb})
corresponds to the case of the broken supersymmetry.
The difference between extended Hamiltonians  (\ref{mathcalLmbr}) and
(\ref{mathcalLm}) is given by the operator
$\mathcal{H}^b_m-\mathcal{L}^e_m=E_{2m-1}+1-2\sigma_3\equiv
\mathcal{I}_m$,
that is an even integral for both supersymmetric systems
$\mathcal{H}^e_m$ and $\mathcal{H}^b_m$. 
This integral does not commute with the supercharges
but generates a rotation in index $\alpha$\,:
$[\mathcal{I}_m,\mathcal{Q}^{e(b),\alpha}_m]=
4i\, \epsilon^{\alpha\beta}\mathcal{Q}^{e(b),\beta}_m$.
\vskip0.1cm

Similarly to what we did in (\ref{AAm}),
one can construct differential operators 
of order $m$,
$\mathbb{A}_{-m}^{-}=A_{-m}^{-}\ldots A_{-1}^{-}$, 
Unlike the  operators $\A_m^-$ and $\A^+_m$, 
they intertwine the system $L_m$ with
the half-harmonic oscillator $L_0$ with an 
additional shift that depends on $m$,
$\A^-_{-m}L_0=(L_m-4m)\A^-_{-m}$,
$\A^+_{-m}L_m=(L_0+4m)\A^+_{-m}.$
This shift also appears  in the relations 
\begin{equation}\label{A-mA-mLiso}
\mathbb{A}_{-m}^{+}\mathbb{A}_{-m}^{-}=\prod_{j=1}^{m}(L_0 
+E_{2j-1})\,,\qquad
\mathbb{A}_{-m}^{-}\mathbb{A}_{-m}^{+}=\prod_{j=1}^{m}(L_m+E_{2j-1}-4m)\,,
\end{equation}
and $L_m\psi_{m,E+4m}=(E+4m)\psi_{m,E+4m}$, where
$\psi_{m,E+4m}=\A_{-m}^-\psi_{E}$.

 \vskip 0.1cm 
 

Due to the shift in the  intertwining relations
generated by 
$A^-_{-m}$ and $A^+_{-m}$ in comparison   
with those for  $A^-_{m}$ and $A^+_{m}$,
the appropriate products of the
operators  from both pairs 
can be used  
to generate the second order ladder 
operators for the isotonic oscillator system $L_m^{\rm iso}$. 
These are 
\begin{equation}
A_m^{-}A_{-m}^{+}=-(a^{-})^2+\frac{m(m+1)}{x^2}\equiv \mathcal{C}_m^{-}\,,
\qquad
\label{11}
A_{-m}^{-}A_{m}^{+}=-(a^{+})^2+\frac{m(m+1)}{x^2}\equiv \mathcal{C}_m^{+}\,,
\end{equation}
where $a^-=\frac{d}{dx}+x$ and $a^+=-\frac{d}{dx}+x$.
The products of the same operators but 
with the changed order
are  the ladder operators for the system $L_{m-1}^{\rm iso}$,
\begin{equation}
A_{-m}^{+}A_m^{-}=-(a^{-})^2+\frac{m(m-1)}{x^2}=
\mathcal{C}_{m-1}^{-}\,,
\qquad
\label{11+}
A_{m}^{+}A_{-m}^{-}=-(a^{+})^2+\frac{m(m-1)}{x^2}=
\mathcal{C}_{m-1}^{+}\,.
\end{equation}
Notice   here that the products
 of the supercharges of the 
different 
extended systems (\ref{mathcalLm})
and (\ref{mathcalLmbr}) generate the
ladder operators for the corresponding  isotonic oscillator 
subsystems\,:
$\mathcal{Q}^{e,1}_m\mathcal{Q}^{b,1}_m=
\text{diag}\,(\mathcal{C}_m^{-},\, \mathcal{C}_{m-1}^{+})$, 
$\mathcal{Q}^{b,1}_m\mathcal{Q}^{e,1}_m=
\text{diag}\,(\mathcal{C}_m^{+},\, \mathcal{C}_{m-1}^{-})$.

In correspondence with the distance 
$\Delta E^{\rm iso}=4$ between 
energy levels in the equidistant spectrum of any 
isotonic oscillator system  $L^{\rm iso}_m$,
the ladder operators  $\mathcal{C}^{+}_m$
and $\mathcal{C}^{-}_m$
satisfy the 
commutation relations 
$[L^{\rm iso}_m,\mathcal{C}^{\pm}_m]=\pm 4\,\mathcal{C}^{\pm}_m$,
$[\mathcal{C}^{-}_m,\mathcal{C}^{+}_m]=8\,L^{\rm iso}_m$
of generators of the $\mathfrak{sl}(2,\R)$  conformal symmetry (\ref{sl2R}).
The kernel of the lowering ladder operator 
is 
\be
\ker\, \mathcal{C}_m^{-}=\text{span}\,\{  (\psi_{m-1,0}^-)^{-1},\psi_{m,0}\  \}\,,
\ee
where 
  $(\psi_{m-1,0}^-)^{-1}$ is 
the non-physical eigenstate of $L^{\rm iso}_{m}$
of eigenvalue $E=-2m+1$, and $\psi_{m,0}$ is the ground state of $L_m^{\rm iso}$ 
of energy 
$E^{\rm iso}_{m,0}=2m+3$.
The kernel of the raising operator $\mathcal{C}_m^{+}$  
is
\be
\ker\, \mathcal{C}_m^{+}=\text{span}\,\{\psi^-_{m,0}\,, (\psi_{m-1,0})^{-1}
 \}\,,
 \ee
where the first and second states are  non-physical eigenstate of $L^{\rm iso}_m$
of eigenvalues $E=-2m-3$ 
 and
$E=2m-1$, respectively.

In correspondence with the fact that the isotonic oscillator $L^{\rm iso}_m$
 has an equidistant spectrum like a half-harmonic  oscillator $L_0^{\rm iso}$, 
 the pair   $\mathcal{C}_m^{+}$ and $\mathcal{C}_m^{-}$
are the spectrum generating operators of  $L^{\rm iso}_m$\,: 
any physical eigenstate of  $L^{\rm iso}_m$ can be produced from  any other of its
physical eigenstate by successive application of these ladder
operators.

Operators $\mathcal{C}_{m-1}^{\pm}$ can be Darboux-dressed  by 
$A_{m}^\pm$ or $A_{-m}^\pm$, but in this way 
we only produce the same ladder operators $\mathcal{C}_m^{+}$ and $\mathcal{C}_m^{-}$ 
multiplied by $(L^{\rm iso}_m + const)$:
$A_{\pm m}^{-}\mathcal{C}_{m-1}^{+}A_{\pm m}^{+}=(L_{m}^{\rm iso}\mp (2m-1))\mathcal{C}_m^{+}$.
Iterating  this procedure,
we can construct  the ladder operators $\mathcal{C}_m^{\pm}$
by Darboux-dressing the ladder operators
$-\mathcal{C}_0^{\pm}=(a^\pm)^2$ of the half-harmonic oscillator $L^{\rm iso}_0$.  
Explicitly,  we have
\be
\mathbb{A}_m^{-}\mathcal{C}_0^{+}\mathbb{A}_m^{+}=\mathbb{A}_m^{-}\mathbb{A}_m^{+}
\mathcal{C}_m^{+}\,,\qquad 
\mathbb{A}_{-m}^{-}\mathcal{C}_0^{+}\mathbb{A}_{-m}^{+}=
\mathcal{C}_m^{+}\mathbb{A}_{-m}^{-}\mathbb{A}_{-m}^{+}\,,
\ee
 where the products  $\mathbb{A}_m^{-}\mathbb{A}_m^{+}$ and 
 $\mathbb{A}_{-m}^{-}\mathbb{A}_{-m}^{+}$  
 are the polynomials of the $L^{\rm iso}_m$ 
 given by equations  (\ref{AmAmgen}) and 
 (\ref{A-mA-mLiso}).

By gluing appropriately different intertwining operators,
we obtain the ladder operators 
$\widetilde{\mathcal{C}}^-_m=\A^-_m\A^+_{-m}$, 
$\widetilde{\mathcal{C}}^+_m=\A^-_{-m}\A^+_{m}$, 
for which 
$[L^{\rm iso}_m,\widetilde{\mathcal{C}}^{\pm}_m]=\pm 4m\,
\widetilde{\mathcal{C}}^{\pm}_m$. 
These ladder operators, however, are not independent
from ${\mathcal{C}}^{\pm}_m$.
Using relations (\ref{11}) 
we  find  that they are reduced just to the degrees of the 
ladder operators $\mathcal{C}^\pm_m$\,:
$\widetilde{\mathcal{C}}^{\pm}_m=(\mathcal{C}^\pm_m)^m$. 
According to  this property,   
the lowering operator  $\widetilde{\mathcal{C}}^{-}_m$
annihilates the  $m$ lowest eigenstates 
of the isotonic oscillator $L^{\rm iso}_m$,
and over each one of these states 
we have an infinite tower of states, where
the operators  $\widetilde{\mathcal{C}}^{\pm}_m$ 
act irreducibly. So, for $m>1$,  the ladder  operators 
$\widetilde{\mathcal{C}}^{\pm}_m$ 
are not spectrum-generating operators, 
and the Hilbert space of the system 
$L^{\rm iso}_m$ is separated into $m$ subspaces 
invariant under their action.
Furthermore we shall see that 
in  the general case of  the rationally extended 
isotonic oscillator systems one can construct
the analogs  of the operators 
 ${\mathcal{C}}^{\pm}_m$ 
 and  $\widetilde{\mathcal{C}}^{\pm}_m$.
 Such analogs  will reproduce some of the 
 described properties of the operators
  ${\mathcal{C}}^{\pm}_m$ 
 and  $\widetilde{\mathcal{C}}^{\pm}_m$,
 but the general picture related to
 ladder operators will be more complicated there.

\vskip0.1cm

In conclusion of this section 
we note that the Hamiltonian operator $L_m^{\rm iso}$ is invariant under the 
change $m\, \,{\rightarrow} -(m+1)$.
If we apply this transformation to the intertwining operators, we obtain
$A_m^{-} \leftrightarrow -A_{-(m+1)}^{+}$ and 
$A_m^{+}  \leftrightarrow  -A_{-(m+1)}^{-}$.
A second application 
reproduces the original operators, i.e. this transformation 
is an involution from the point of view of the two families  of 
the Darboux transformations that we 
discussed.
The ladders operators $\mathcal{C}_{m}^\pm$ are invariant under
 this transformation.
On the other hand, the stationary Schr\"odinger equation for the isotonic oscillator
$L_m^{\rm iso}\psi(x)=E\psi(x)$ is invariant under the transformation $x 
 \, \,{\rightarrow}\,\, ix$ accompanied by a change  $E \, \, {\rightarrow}  -E$. 
Under this  transformation 
$A_m^{\pm} \, \,{\rightarrow} -iA_{-m}^{\pm}$,  
$A_{-m}^{\pm} \, \,{\rightarrow} -iA_{m}^{\pm}$,  
and 
$\mathcal{C}^-_m\, \,{\rightarrow} \, -\mathcal{C}^+_m$, 
$\mathcal{C}^+_m\, \,{\rightarrow} \, -\mathcal{C}^-_m$.
This transformation of the ladder operators reflects the fact 
that the invariance of the stationary Schr\"odinger equation
for isotonic oscillator also
requires the indicated change of the energy sign. 
The double application of this transformation
to the intertwining operators provokes just 
their multiplication by the phase $-1$, i.e. it also is
the involution.

\section{Dual schemes and mirror diagrams}\label{SectionDual}

Here, we describe the construction of dual schemes of  DC
transformations which  can conveniently be presented
by a mirror diagram.  Via a ``charge conjugation" procedure,  
the mirror diagram allows us, in turn, to  easily  reconstruct a dual partner if  
one of the two  schemes is given. 
The dual schemes and mirror diagrams 
 will play a key role in the constructions of the next two sections.

In the previous section we saw that the isotonic oscillator systems 
can be generated  from the half-harmonic oscillator $L^{\rm iso}_0=L_0$  
by applying to it  DC transformations 
which use as the sets of  seed states either only physical eigenstates 
of $L_0$,  
or only its specific non-physical eigenstates that are 
obtained from the physical
states by the spatial  Wick rotation. 
We shall show now that these two schemes can be related  one to  another
via  a chain of intermediate DC  schemes 
which are based on  mixed sets of the physical and non-physical
eigenstates.  This will be done by us 
for the pairs of DC transformations 
of a more general form which will be used in subsequent sections
for construction of the rational  extensions 
of the isotonic
oscillator systems. 
To this aim we first note that the indicated above 
sets of physical and non-physical eigenstates of the
half-harmonic oscillator $L_0$ are also the eigenstates 
of the original quantum harmonic oscillator  $L^{\rm osc}$.
Then, let us consider the DC transformation generated
by a set of  non-physical eigenstates 
$(-n_m,\ldots,-n_1)$ of $L^{\rm osc}$. We assume here
that $n_m>\ldots>n_1>0$, and do not preoccupy about
zeros of the corresponding Wronskian 
$W(-n_m,\ldots,-n_1)$.  It is rather natural to call such a scheme
``negative".  The scheme which will involve in the
set of  seed states $(n_1',\ldots, n'_l)$, $0<n'_1<\ldots <n'_l$, 
only physical eigenstates 
of $L^{\rm osc}$ we call ``positive".
Using the properties of the Wronskians described in Appendix \ref{subSecDC},
we can write  the relation 
$W(-n_m,\ldots,-n_1)=W(-0,\widetilde{\,-0\,},-n_m,\ldots,-n_1)$. 
As before,  an equality between the corresponding 
Wronskians implies here  the equality modulo a nonzero constant multiplier 
which  has no effect on the corresponding DC 
transformations.  Since the state $(-0)=\psi^-_0$ generates,
according to (\ref{defApsi}),  the first order differential operator 
$-a^+$, and $a^+\widetilde{\psi^-_0}=\psi_0$ \cite{CarPly1},
we obtain  for $W(-n_m,\ldots,-n_1)$ the equivalent representations
\be
\label{reglamagica}
e^{\frac{x^2}{2}}W(a^{+}(\widetilde{\,-0\,}),a^{+}(-n_m),\ldots,a^{+}(-n_1)) 
= e^{\frac{x^2}{2}}W(0,-(n_m-1),\ldots,-(n_1-1))\,.
\ee
Having in mind that the DC transformation 
generates an additional  potential term  $-2(\ln W)''$, we see
that the ``negative" scheme $(-n_m,\ldots,-n_1)$ and the 
scheme  $(0,-(n_m-1),\ldots,-(n_1-1))$, that 
involves a mixed set of physical and nonphysical 
eigenstates of $L^{\rm osc}$, 
generate the same quantum Hamiltonian up 
to a constant relative shift equal to $2$.
Let us assume that $(n_1-1)>0$.  In this case the last
argument of the Wronskian in the equality in (\ref{reglamagica}) 
will be a non-physical eigenstate of $L^{\rm osc}$ different from $(-0)$.
Then  
we have  a relation 
\be
W(0,-(n_m-1),\ldots,-(n_1-1))=W(-0,\widetilde{\,-0\,},0,-(n_m-1),\ldots,-(n_1-1))\,.
\ee
The chain of relations of the form (\ref{reglamagica}) 
can be iterated,  and  we obtain
\be
W(-n_m,\ldots,-n_1)
= e^{x^2}W(0,1,-(n_m-2),\ldots,-(n_1-2))\,. 
\ee
In this second step the power 
of the exponent is multiplied by two and now we have 
some new mixed scheme that includes two physical eigenstates 
of $L^{\rm osc}$. 
If $n_{1}-2>0$, we repeat again the  procedure till the moment 
when in the last argument of the Wronskian we obtain the 
non-physical eigenstate $(-0)$.
When this happens, 
we move this state  to the first  position in the Wronskian,
and use  the relation $W(-0,0,\ldots,-(n_1-k))=e^{x^2/2}W(1,\ldots,-(n_1-(k+1)))$.
No matter of whether the last argument in the Wronskian is a 
non-physical state different from $(-0)$ or equal to it,
we apply here one of the two
corresponding algorithms described above,
and stop the chain of the transformations and equalities
when all the arguments of the Wronskian will be physical eigenstates
of the harmonic oscillator.  
In such a way we arrive finally at the relation 
of the form
\be\label{genDual}
W(\underbrace{-n_m,\ldots,-n_1}_\text{\,\,\,\,\,$n_-$})=e^{(n_m+1)x^2/2}
W(\underbrace{n'_1,\ldots,n_m'=n_m}_\text{$\,\,\,\,\,n_+$})\,,
\ee
where $0<n'_1<\ldots<n'_m=n_m$.
 This relation means that 
the ``negative" scheme $(-n_m,\ldots,-n_1)$ 
with $n_-$ seed states is dual 
to the ``positive" scheme $(n'_1,\ldots,n'_m=n_m)$
with $n_+=n_m+1-n_-$ seed states,
and that the corresponding DC transformations
generate the same Hamiltonian operator  modulo 
a relative constant shift $2(n_m+1)$. 

In the simplest case we obtain in the described
 way the chain of relations 
\begin{eqnarray}
(-1)&=&W(-0,\widetilde{\,-0\,},-1)=e^{x^2/2}W(a^+(\widetilde{\,-0\,}),a^+(-1))\nonumber\\
&=&
e^{x^2/2}W(0,-0)=e^{x^2/2}W(-0,0)=e^{x^2}W(a^+(0))=e^{x^2}\cdot (1)\,.
\end{eqnarray}
This is generalized to a relation
\be
W(-1,-3,\ldots,-(2n_m-1),-(2n_m+1)) =e^{(n_m+1)x^2}W(1,3,\ldots,2n_m-1,2n_m+1)
\ee
which corresponds to the pair of relations 
(\ref{LmWron}) and (\ref{LmWron-4m}).
Here   we also used the property  that the permutation of 
arguments in the Wronskians has no effect for the 
corresponding DC transformation.
Another useful relation is 
$(-n)=W(-n)=e^{(n+1)x^2/2}W(1,2,\ldots,n)$.
This relation and the corresponding duality of the schemes 
$(-n)$ and $(1,2,\ldots,n)$, which we indicate by  
$(-n)\sim(1,2,\ldots,n)$, corresponds to 
isospectral deformation of the isotonic 
oscillator $L^{\rm iso}_1$ and will be discussed in the 
next section.
The particular relation 
\begin{equation}
\label{esquemajemplo}
W(-2,-3,-4,-5,-8,-9,-11)= e^{6x^2}W(1,4,5,10,11)
\end{equation}   
and the duality of the corresponding schemes 
$(-2,-3,-4,-5,-8,-9,-11)\sim (1,4,5,10,11)$
will be used by us in an example of Section \ref{SecGapped}.

We could start, instead,  from some 
``positive" scheme. Then  we  introduce into the set of arguments
of the Wronskian the pair of the states 
$0$ and $\widetilde{0}$, and  use the relations 
$A^-_0=a^-$, $a^-(\widetilde{0})=(-0)$,
$a^- (n)=(n-1)$, $W(0,n,\ldots)=e^{-x^2/2}
W(a^-(n),\ldots)$. The final result will be 
presented by relation of the same form (\ref{genDual}).

\begin{figure}[H]
\begin{center}
\includegraphics[scale=0.16]{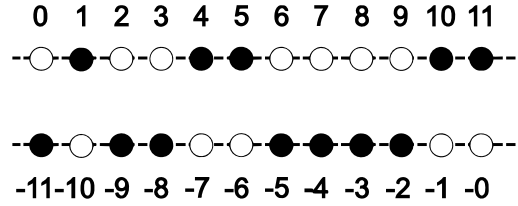} 
\caption{The mirror diagram for  the dual schemes in (\ref{esquemajemplo}).} 
\label{Figure1}
\end{center} 
\end{figure}

The described procedure of   construction
of the dual schemes can be summarized and presented 
in the form of the corresponding mirror diagrams.
Particularly, the duality of the schemes corresponding 
to relation (\ref{esquemajemplo}) is presented
by the mirror diagram shown on  Figure  
\ref{Figure1}.
\vskip-0.5cm
The mirror diagram and the corresponding 
dual schemes are constructed in the following way.
Let us take a  ``positive" scheme given
by the set of the seed states $(n_1,\ldots,n_m)$ with
$0<n_1<\ldots<n_m$.
This scheme we present by the upper horizontal line,
where we mark by
unfilled, white circles the ground state $(0)$ of $L^{\rm osc}$ and all 
physical eigenstates with positive index less then $n_m$ which 
do not belong to the set of numbers $n_1,\ldots, n_m$. 
The states $n_1,\ldots, n_m$, which are selected as the 
seed states,  we mark by filled, black circles.
The dual, ``negative" scheme is presented by the lower horizontal line,
where the  non-physical states $(-n)$ which are used or  not
used in the scheme  as the seed states are marked similarly 
by black and white circles, respectively.
The circles in the ``negative" scheme are obtained from those
in the ``positive"  scheme by a kind of a ``charge conjugation" procedure\,:
below  a  white (black) circle in the upper line we put
a black (white) circle in the lower line. 
Then we assign the set of consecutive 
numbers $(-n_m, -n_m+1,\ldots,-1, -0)$
to the circles in the lower line by putting $(-n_m)$ below
the state $(0)$ in the upper line and $(-0)$ below the 
state $n_m$. If, instead,  the ``negative"  scheme is given, 
we present it  by the lower line, 
and reconstruct the upper line of the dual, 
``positive" scheme in the obvious way
by inverting the described procedure of the ``charge conjugation".

Let us denote the number of black circles in the upper 
line by $n_+$. Then the number of black circles in the
lower line will be $n_-=n_m+1-n_+$.
Remembering that each physical eigenstate
$(n)$ carries the exponential factor $e^{-x^2/2}$,
while non-physical state $(-n)$ carries the factor
$e^{x^2/2}$, we  obtain  the relation between Wronskians
of the dual schemes in the form $W(-n_m,\ldots,-n_1)e^{-(n_m+1-n_+)x^2/2}=
W(n_1',\ldots,n_m)e^{n_+x^2/2}$,
that  corresponds to 
(\ref{genDual}).

In order the dual schemes would 
produce the potentials
which will describe  non-singular
rational deformations
of the conformal mechanics  systems, the seed states 
have to be chosen  
in such a way that the corresponding Wronskians 
will have no zeros on the half-line $(0,\infty)$. 
In the next two sections,  we describe how 
non-singular dual  schemes can 
be designed, and use them 
for the construction of ladder operators for corresponding 
rationally deformed isotonic oscillator systems.

\section{Completely  isospectral rational deformations}\label{SectionIso}

In Section \ref{SectionIsotonic}, we noticed  that the 
selection 
of a set of $m$ 
non-physical eigenstates $\psi_{2l+1}^{-}$, 
$l=0,1,...,m-1$, 
of the half-oscillator 
$L^{\rm osc}_0$ 
as the  seed states for the DC transformation
produces the isotonic 
oscillator system  $L_m^{\rm iso}$.   
In this section we show that the choice of the 
same set of non-physical eigenstates as the seed states 
but taken with gaps in the set of values
of the parameter $l$ generates completely isospectral
deformations of the isotonic oscillator systems.
In spite of the complete isospectrality
with the corresponding isotonic oscillator of index $m$,  each  such 
a system will be characterized by peculiar lowering and raising 
ladder operators that will be spectrum-generating 
operators of differential order  higher than two.
For each  completely isospectral deformation of 
conformal mechanics  system we also construct an additional 
pair of  ladder operators whose analogs
will play an important role  later  for non-isospectral deformations 
of the isotonic oscillator systems.

The generation   
of 
completely isospectral rational deformations 
of the isotonic  oscillator  system $L^{\rm iso}_m$ is based 
on the relation
\begin{equation}
\label{reio1}
W(\psi_{2l_1+1}^-,\psi_{2l_2+1}^-,...,\psi_{2l_m+1}^-)=
e^{mx^2/2}x^{m(m+1)/2}f_{(-(2l_1+1),\ldots,-(2l_{m}+1))}(x)\,,
\end{equation}
where $f_{(-(2l_1+1),\ldots,-(2l_{m}+1))}(x)$ 
is an even  polynomial function with no real roots, and 
$0\leq l_1<l_2<\ldots<l_m$, see Appendix \ref{AppB}.
In  the case of the choice of the 
non-physical eigenstates with indices given
by a consecutive set  $l_1=0,\, l_2=1,\ldots,\,l_m=m-1$
without any gap, polynomial function $f$ is reduced just to
a nonzero constant, and Wronskian (\ref{reio1})  
generates by means of relation
(\ref{LmWron-4m}) the isotonic oscillator system
$L^{\rm iso}_m$ shifted for a constant $-2m$. 
The appearance of any gap in the set of values 
$0\leq l_1<l_2<\ldots<l_m$ gives rise to a nontrivial 
polynomial $f(x)$ of even order that generates 
a  completely isospectral deformation of the 
shifted isotonic oscillator system 
$L^{\rm iso}_m-2m$. 
The corresponding Schr\"odinger operator 
generated by the DC transformation 
has the form  
\be
L_{(-(2l_1+1),\ldots,-(2l_{m}+1))} 
= 
L_m^{\rm iso}-2(\ln f_{(-(2l_1+1),\ldots,-(2l_{m}+1))})''-2m\,, 
\label{reio2} 
\ee 
and describes  the isotonic Hamiltonian  
extended with a nonsingular on the 
half-line $x>0$  rational term. The nonsingular character of 
the additional rational term is related to the nature of the polynomial function 
$f_{(-(2l_1+1),\ldots,-(2l_{m}+1))}(x)$ described above. 
In (\ref{reio2}),  the lower index in the notation
of the Hamiltonian operator indicates what  (non-physical
in this case)
eigenstates of the quantum harmonic oscillator are used 
as the seed states in the corresponding DC trasformation.
A similar notation will be used below  for the 
intertwining operators  being generators of the 
corresponding Dabroux-Crum transformations.
The complete isospectrality of (\ref{reio2}) to the shifted isotonic oscillator 
$L^{\rm iso}_m$
takes place  because no new 
normalizable states are aggregated in this new system and its spectrum, up to a global shift,
 has to coincide with the spectrum of $L_0$.

For each fixed integer 
 $m\geq 1$ we  have 
an infinite number
of possibilities to choose $m$  seed states in the form of the 
set of non-physical eigenstates  of
$L^{\rm osc}$.  
In this way we can construct an infinite family of  the deformed isotonic oscillator 
systems that will  be completely isospectral to the 
shifted isotonic oscillator system $L^{\rm iso}_m-2m$.
For each such a  system, there is another
dual   DC transformation
to generate it. The existence of  the dual DC schemes 
is essential for the construction of the ladder operators 
for the rationally extended isotonic oscillator systems. 

As an example,  
consider the case of the simplest 
deformed 
isotonic oscillator system generated via the 
Darboux transformation
based on the non-physical eigenstate 
$\psi_{3}^{-}=(2x^3+3x)e^{x^2/2}$
of the half-harmonic oscillator $L^{\rm iso}_0$. 
In this case $f_{(-3)}(x)=2x^2+3$,
and Eq. (\ref{reio2}) reduces to 
\begin{eqnarray}
\label{reio3}
L_{(-3)} 
=L_1^{\rm iso} -2 +8\frac{2 x^2-3}{(2 x^2+3)^2}\,.
\end{eqnarray}
By the method of the mirror diagram, 
we find that up to a constant shift,
the system can be generated alternatively by
the  DC 
transformation based on the set  $\{\psi_1, \psi_2, \psi_3\}$ 
of
physical eigenstates
of the quantum harmonic oscillator, 
$L_{(1,2,3)}=
L_{(-3)}+8$.
The states $\psi_1$ and  $\psi_3$ here are the two lowest
physical eigenstates of the half-harmonic oscillator,
whereas the state $\psi_2$ is a  non-physical
eigenstate of $L_0=L_0^{\rm iso}$ that does not satisfy 
the boundary condition at $x=0$.

Though the schemes  based  
on the two  different sets of  seed states $(-3)$ and $(1,2,3)$
generate, up to a constant shift, the same system, 
the corresponding associated intertwining operators
are essentially different. 
The intertwiners 
\be\label{A+-(-3)}
A_{(-3)}^{-}=\psi^{-}_3\frac{d}{dx}\frac{1}{\psi^{-}_3}
=A_{-1}^- -\frac{4x}{2x^2+3}\,,\qquad
A_{(-3)}^{+}=-\frac{1}{\psi^{-}_3}\frac{d}{dx}\psi^{-}_3
=A_{-1}^+-\frac{4x}{2x^2+3}\,,
\ee
are the extensions of the first order differential operators 
$A^\pm_{-1}$ of the form  (\ref{A-m}) with $m=1$.
The appearance of the operators $A^\pm_{-1}$ in the structure of 
the DC generators $A_{(-3)}^{\pm}$ in 
(\ref{A+-(-3)})
reflects the fact that the system (\ref{reio3})
represents a rational extension 
of the isotonic  oscillator system $L^{\rm iso}_1$.
The additional rational term
$-{4x}/{(2x^2+3)}$ in (\ref{A+-(-3)}) is regular on
all 
the 
half-axis $x>0$ and tends  to zero
for $x\rightarrow 0$ and $x\rightarrow +\infty$.
We have the factorization relations 
$A_{(-3)}^{+}A_{(-3)}^{-}=L_0+7$, 
$A_{(-3)}^{-}A_{(-3)}^{+}=L_{(-3)}+7=L_{(1,2,3)}-1$. 
In correspondence with them, 
$A_{(-3)}^{-}$ intertwines the Hamiltonian operators 
$L_0$ and $L_{(-3)}$,
\be
A_{(-3)}^{-}L_0=L_{(-3)}A_{(-3)}^{-}=(L_{(1,2,3)}-2\Delta E^{\rm iso})A_{(-3)}^{-}\,, 
\ee
where the constant $2\Delta E^{\rm iso}=8$
will reveal  itself in the properties of the ladder operators.
The intertwining relation for $A_{(-3)}^{+}$ is obtained 
by Hermitian conjugation.  

The systems $L_0$ and $L_{(1,2,3)}$ are also intertwined 
by the third order operators $\A^{-}_{(1,2,3)}$
and $\A^{+}_{(1,2,3)}=(\A^{-}_{(1,2,3)})^\dagger$, where 
the operator $\A^{-}_{(1,2,3)}$ is uniquely 
 specified  by 
 its kernel\,: $\ker \A^{-}_{(1,2,3)}=\text{span}\,\{\psi_1,\psi_2,\psi_3\}$.
 Notice  that 
 the factorized form of the operators $\A^{\pm}_{(1,2,3)}$
 is not unique, but  it is not important for us at the moment.
 In such notations, the operators $\A^-_m$ and $\A^-_{-m}$,
 that we discussed before for generation of isotonic oscillator systems,
take,  in correspondence with  Eqs. (\ref{LmWron})
and (\ref{LmWron-4m}), the form 
$\A^-_m=\A_{(1,3,\ldots,2m-1)}^-$ and 
$\A^-_{-m}=\A_{(-1,-3,\ldots,-(2m-1))}^-$. 
We have the intertwining relation 
$\A^{-}_{(1,2,3)}L_0=L_{(1,2,3)}\A^{-}_{(1,2,3)}=(L_{(-3)}+8)\A^{-}_{(1,2,3)}$, 
and the conjugate relation for $\A^{+}_{(1,2,3)}$.

Analogously to the case of the isotonic oscillator systems,
we can construct the ladder operators for the system 
$L_{(-3)}=L_{(1,2,3)}-8$ by Darboux-dressing of the
ladder operators of the half-harmonic oscillator.
Doing this with the first order intertwining operators, we 
obtain
\begin{equation}
\label{reio7}
\mathcal{A}^{\pm}=A_{(-3)}^{-}(a^{\pm})^{2}A_{(-3)}^{+}\,.
\end{equation}  
These  operators together with the Hamiltonian $L_{(-3)}$
generate a nonlinear deformation of the conformal symmetry 
given by the commutation relations
\be
[L_{(-3)},\mathcal{A}^{\pm}]=\pm 4  
\mathcal{A}^{\pm}, \qquad
[\mathcal{A}^-,\mathcal{A}^+]=16\left(L_{(-3)}+3\right)\left(L_{(-3)}+7\right)\left(L_{(-3)}+{1}/{2}\right).
\ee

The roots of the  fourth order polynomial 
in the relation 
\be
\mathcal{A}^{+}\mathcal{A}^{-}= 
(L_{(-3)}+7)(L_{(-3)}+3)(L_{(-3)}-1)(L_{(-3)}-3)
\ee
 correspond to  eigenstates of
$L_{(-3)}$  which belong to the 
kernel of the lowering  operator, 
\be
\ker\mathcal{A}^{-}=\text{span}\,\{ A_{(-3)}^{-}
\widetilde{\psi_{3}^{-}},\,A_{(-3)}^{-}\psi_1^{-}\,,
A_{(-3)}^{-}\psi_0,\,
A_{(-3)}^{-}\psi_1 
\}.
\ee
The last state $\,A_{(-3)}^{-}\psi_1=\A_{(1,2,3)}^-\psi_5$ describes here the ground state
of $L_{(-3)}$ of eigenvalue  $E=3$.
In comparison with the second order ladder operator
$(a^{-})^{2}$ of the half-harmonic oscillator, the
fourth order lowering operator $\mathcal{A}^{-}$
contains besides the ground state three non-physical
eigenstates of the  Hamiltonian $L_{(-3)}$
instead of one. 
The roots in the  alternative product 
\be
\mathcal{A}^{-}\mathcal{A}^{+}=(L_{(-3)}+11)(L_{(-3)}+7)(L_{(-3)}+3)(L_{(-3)}+1)
\ee
correspond to 
eigenvalues of the eigenstates of $L_{(-3)}$
which appear in the kernel of the raising ladder operator,
\be
\ker\mathcal{A}^{+}=\text{span}\{ A_{(-3)}^{-}\psi_5^{-},\,
A_{(-3)}^{-}\widetilde{\psi_{3}^{-}},\, 
A_{(-3)}^{-}\psi_1^{-},\,A_{(-3)}^{-}\psi_0^{-}\}\,.
\ee 
All the states in this kernel are non-physical.
In correspondence with the described properties of 
the ladder operators (\ref{reio7}) they
are the spectrum-generating operators for the
system $L_{(-3)}$\,:
acting by them on any physical eigenstate of $L_{(-3)}$,
we can generate any other physical eigenstate.
The kernels of the ladder operators 
contain here the same \emph{non-physical} eigenstate $A_{(-3)}^{-}\widetilde{\psi_{3}^{-}}=
A_{(-3)}^{-}\psi_1^{-}$. Below we shall see that in the
case of non-isospectral rational deformations of the isotonic 
oscillator systems the kernels of analogs 
of such lowering and raising ladder operators contain 
some common \emph{physical} eigenstates.

In a similar way, one can construct 
the ladder operators for  $L_{(-3)}$  
via Darboux-dressing of  the ladder operators $(a^\pm)^2$ 
of $L_0$ 
by the third order intertwining operators,
$\mathcal{B}^{\pm}=\A^{-}_{(1,2,3)}(a^{\pm})^2\A^{+}_{(1,2,3)}$, 
$[L_{(-3)},\mathcal{B}^{\pm}]=\pm 4\mathcal{B}^{\pm}$.
However, these  differential operators of order $8$ are not independent 
and reduce to the fourth order ladder operators 
(\ref{reio7}) multiplied by the second order polynomials in 
the Hamiltonian,
\be
\mathcal{B}^{-}=\mathcal{A}^{-}(L_{(-3)}+1)(L_{(-3)}+5)\qquad
\text{and} \qquad
\mathcal{B}^{+}=(\mathcal{B}^{-})^\dagger\,. 
\ee
In correspondence with these relations, the
operator $\mathcal{B}^{-}$ annihilates four non-physical 
eigenstates 
$A^{-}_{(-3)}\psi_0^{-}$,
$A^{-}_{(-3)}\widetilde{\psi_0^{-}}$,
$A^{-}_{(-3)}\psi_2^{-}$,
and
$A^{-}_{(-3)}\widetilde{\psi_2^{-}}$
of  the quantum Hamiltonian $L_{(-3)}$
in addition to those states 
which constitute the kernel of the ladder operator 
$\mathcal{A}^{-}$. Analogously, 
the kernel of $\mathcal{B}^{+}$
is composed from non-physical states of the 
kernel of  the operator $\mathcal{A}^{+}$ 
extended by four non-physical eigenstates
$A_{(-3)}^{-}\psi_4^{-}$,
$A_{(-3)}^{-}\widetilde{\psi_{4}^{-}}$,
$A_{(-3)}^{-}\psi_2^{-}$, and
$A_{(-3)}^{-}\widetilde{\psi_{2}^{-}}$
of the Hamiltonian $L_{(-3)}$.
Since all the additional states from the kernels 
of the operators $\mathcal{B}^{-}$ and $\mathcal{B}^{+}$
in comparison  with the kernels of the ladder
operators $\mathcal{A}^{-}$ and $\mathcal{A}^{+}$
are non-physical, the  
$\mathcal{B}^{-}$ and $\mathcal{B}^{+}$
are also the spectrum-generating operators 
of the system $L_{(-3)}$.
So, in this case one can conclude that the order eight 
ladder operators $\mathcal{B}^{\pm}$ are, in fact, equivalent 
to the fourth order ladder operators 
(\ref{reio7}) when they act on physical eigenstates. 
We shall see  later that in the case of the 
non-isospectral rational extensions 
of the isotonic oscillator systems different 
Darboux-dressing procedures produce 
non-equivalent pairs of the ladder operators
which will reflect different properties
of the corresponding quantum systems.

Consider the construction 
of yet another pair of ladder operators  for the same system
$L_{(-3)}$  by Darboux-dressing the ladder operators
of the isotonic oscillator $L_2^{\rm iso}$.
For this we first note that the iterative 
nature of the DC transformation
allows us to present the third order
intertwining operators $\A_{(1,2,3)}^{\pm}$
in the following particular factorized form\,:
\begin{equation}\label{A123A2A}
\A_{(1,2,3)}^{-}=A_{(2)}^{(1,3)-}\A_{(1,3)}^- \,,\qquad
\A_{(1,2,3)}^{+}= \A_{(1,3)}^+A_{(2)}^{(1,3)+}\,.
\end{equation} 
Here the second order differential operators 
$\A_{(1,3)}^-=\A_2^{-}=A^-_2A^-_1$ and $\A_{(1,3)}^+=(\A_{(1,3)}^-)^\dagger=\A_2^{+}$ 
correspond exactly to the particular 
case $m=2$ of the operators (\ref{AAm})
that intertwine the half-harmonic oscillator
$L_0$ and the shifted isotonic oscillator system $L_2$.
The first order differential operators
$A_{(2)}^{(1,3)-}$ and $A_{(2)}^{(1,3)+}$
are given by relations
$A_{(2)}^{(1,3)-}=\left(\A_2^{-}\psi_2\right)
\frac{d}{dx}\left( 
{\A_2^{-}\psi_2}
\right)^{-1}$, 
$A_{(2)}^{(1,3)+}=(A_{(2)}^{(1,3)-})^{\dagger}$. 
Representation (\ref{A123A2A}) can be related  to
the 
Wronskian identity
$W(1,2,3)=-W(1,3,2)$,  
and corresponds to a general rule of association 
of the $n$th order intertwining operators
with the DC transformation of order $n$ 
described in Appendix \ref{subSecDC}.
The upper index  in the  operator
$A_{(2)}^{(1,3)-}$ 
indicates the seed states 
of the harmonic oscillator
we use for the construction of the 
second order operators $\A^\pm_2$,
and the lower index indicates the eigenstate 
$\psi_2$ of $L^{\rm osc}$ 
to which we apply the  operator $\A^-_2$
to generate finally the  operators $A_{(2)}^{(1,3)-}$ and 
$A_{(2)}^{(1,3)+}$.

As $\A_2^{-}\psi_2=x^{-2}(3 + 2 x^2)e^{-x^2/2}$ 
is a (non-physical) eigenstate of $L_2$,
which can be obtained
by applying the symmetry transformation 
$m \rightarrow -m-1$ to the first excited state
of $L_2$ by means of  Eq. (\ref{LnHn}),
the operators $A_{(2)}^{(1,3)-}$ and 
$A_{(2)}^{(1,3)+}$ factorize the appropriately 
shifted Hamiltonian operators $L_2$ and $L_{(1,2,3)}$,
\be
A_{(2)}^{(1,3)+}A_{(2)}^{(1,3)-}=L_2-5\,, \qquad
A_{(2)}^{(1,3)-}A_{(2)}^{(1,3)+}=L_{(1,2,3)}-5\,, 
\ee
and intertwine them, 
\be
A_{(2)}^{(1,3)-}L_2=L_{(1,2,3)}A_{(2)}^{(1,3)-}\,,
\qquad
A_{(2)}^{(1,3)+}L_{(1,2,3)}=L_2A_{(2)}^{(1,3)+}\,.
\ee
Based on these properties and relations,
we can construct the ladder operators for 
the Hamiltonian operator  $L_{(1,2,3)}=L_{(-3)}+8$
by Darboux-dressing the ladder operators
$\mathcal{C}^\pm_2$ of the isotonic 
oscillator  system $L_2$,
$\widetilde{\mathcal{B}}^{\pm}=A_{(2)}^{(1,3)-} \mathcal{C}_2^{\pm}  A_{(2)}^{(1,3)+}$. 
A simple calculation 
of the products $\widetilde{\mathcal{B}}^{-}\widetilde{\mathcal{B}}^{+}$ 
and $\widetilde{\mathcal{B}}^{+}\widetilde{\mathcal{B}}^{-}$ 
shows that  
the kernels of the ladder operators $\tilde{\mathcal{B}}^-$
and  $\tilde{\mathcal{B}}^+$
 coincide with the kernels  of the operators
 ${\mathcal{A}}^-$
and  ${\mathcal{A}}^+$, 
 and  we find that 
 the operators $\widetilde{\mathcal{B}}^{\pm}$  are the same
 ladder operators  as (\ref{reio7})\,: $\widetilde{\mathcal{B}}^{\pm}=-{\mathcal{A}}^{\pm}$.

Intuitively, one can expect  that for each system 
completely isospectral to the 
half-harmonic oscillator $L_0$, all the ladder operators
 constructed 
by Darboux-dressing  do the same work and that
 they are related between themselves
by a simple multiplication with a polynomial 
in the corresponding  Hamiltonian. 
This regularity appears for both the isotonic 
oscillator systems and for their isospectral 
rational extensions. However,  having in mind 
the analogy with the rationally extended  quantum harmonic 
oscillator systems 
\cite{CarPly2},  
we cannot expect the  same picture  in  non-isospectral 
deformations of the isotonic oscillator systems
having gaps in the  spectra.

As 
the first and third
order operators 
$A_{(-3)}^{\pm} $ and $\A_{(1,2,3)}^{\pm}$
intertwine the half-harmonic oscillator with the
system $L_{(-3)}$ with a nonzero relative 
shift,
we can 
construct 
yet another pair of the ladder operators
for the quantum system $L_{(-3)}$,
 \begin{equation}\label{C+-iso}
 \mathcal{C}^{-}=\mathbb{A}_{(1,2,3)}^{-}A_{(-3)}^{+}\,, 
 \qquad
 \mathcal{C}^{+}=A_{(-3)}^{-}\mathbb{A}_{(1,2,3)}^{+} \,.
\end{equation}
These   forth order differential operators
have a structure similar to the factorized structure 
of the 
ladder operators $\mathcal{C}_m^{\pm}$
for isotonic 
oscillator systems  that we considered in Section 
\ref{SectionIsotonic}.
Unlike the ladder operators $\mathcal{A}^\pm$, 
$\mathcal{B}^\pm$ and $\widetilde{\mathcal{B}}^\pm$,
the $\mathcal{C}^{\pm}$
satisfy the commutation relations 
$[L_{(-3)},\mathcal{C}^{\pm}]=\pm 2\Delta E^{\rm iso}\mathcal{C}^{\pm}$, 
where the coefficient $2\Delta E^{\rm iso}=8$ corresponds 
to the relative shift between the operators 
 $L_{(1,2,3)}$ and $L_{(-3)}$, and is equal to the double 
distance between the energy levels in the spectrum of the 
rationally extended isotonic system $L_{(-3)}$.
They  generate the following polynomial deformation
 of the conformal symmetry\,:
 \be
 [L_{(-3)},\mathcal{C}^{\pm}]=\pm 8\,\mathcal{C}^{\pm}\,,\qquad
 [\mathcal{C}^-,\mathcal{C}^+]=32\left(L^3_{(-3)}+6L^2_{(-3)}-L_{(-3)}+30\right)\,.
 \ee
The kernel of the lowering ladder operator  
 is
 \be
\ker\mathcal{C}^{-}=\text{span}\,\{(\psi_{(-3)}^{-})^{-1},\,A_{(-3)}^{-}\psi_1,\,A_{(-3)}^{-}\psi_2,\,A_{(-3)}^{-}\psi_3
\}\,.
\ee
Here
 $A_{(-3)}^{-}\psi_1=\A_{(1,2,3)}^-\psi_5$ and $A_{(-3)}^-\psi_3=\A_{(1,2,3)}^-\psi_7$
  are the ground  and the first exited states of $L_{(-3)}$. 
 All the states in the kernel of 
the raising ladder operator are non-physical\,:
\be
\ker \mathcal{C}^{+}=\text{span}\,\{A_{(-3)}^{-}\psi_7^{-},\,A_{(-3)}^{-}\psi_2^{-},\,A_{(-3)}^{-}
\psi_1^{-},\,A_{(-3)}^{-}\psi_0^{-}\}\,.
\ee
 As a result, the space of states
 of  $L_{(-3)}$ is separated into  two subspaces, on each of which
 the ladder operators $\mathcal{C}^{+}$ and $\mathcal{C}^{-}$ 
 act irreducibly. One subspace is spanned  by
 the ground state  
 and the corresponding excited eigenstates, 
 $A_{(-3)}^{-}\psi_{4l+1}=
  (\mathcal{C}^{+})^l(A_{(-3)}^{-}\psi_1)=
  (\mathcal{A}^+)^{2l}(A_{(-3)}^{-}\psi_1)$, 
  $l=0,1,\ldots$.
Another subspace corresponds 
to the  first excited state and the  
infinite tower of the states produced from  it by application of the raising operator 
$\mathcal{C}^{+}$,
 $A_{(-3)}^{-}\psi_{4l+3}=
  (\mathcal{C}^{+})^l(A_{(-3)}^{-}\psi_3)=
  (\mathcal{A}^+)^{2l}(A_{(-3)}^{-}\psi_3)$, 
  $l=0,1,\ldots$.
  Here, as in other relations, we imply equalities modulo
  nonzero multliplicative factors.
  The ladder operators 
 $\mathcal{C}^{\pm}$,  unlike $\mathcal{A}^{\pm}$,
 are therefore not spectrum-generating operators  for the system 
 $L_{(-3)}$. 
 {}Notice that from the point of view of the basic properties 
 of the ladder operators  $\mathcal{C}^{\pm}$,
 they are similar to the operators 
 $(\mathcal{C}_0^\pm)^2=(a^\pm)^4$ 
 in the case of the half-harmonic oscillator
 $L_0$, or to the operators  $\widetilde{\mathcal{C}}_2^\pm$
 for isotonic oscillator $L_2^{\rm iso}$.
 The essential difference here, however,   is that
 the ladder operators
 $\mathcal{C}^{\pm}$ are independent from 
  the spectrum-generating
 ladder operators  $\mathcal{A}^{\pm}$
 and have the same  differential order equal to four.
 We shall see  
 that for non-isospectral
 rational extensions of the isotonic oscillator systems 
 the direct analogs of the operators  $\mathcal{C}^{\pm}$
 will constitute an inseparable part of the set
 of the spectrum-generating operators.

\vskip.1cm

 Let us also note here yet another possibility 
 to generate the same system $L_{(-3)}$.
 We first can choose the neighbour excited eigenstates $\psi_2$ 
 and $\psi_3$ of the quantum harmonic oscillator
 as the seed states for the DC transformation.
 In this way we generate the following  rational extension of the
 quantum harmonic oscillator\,: 
\be
L_{(2,3)}= 
-\frac{d^2}{dx^2} + x^2 - \frac{16 x^2 (4 x^4-9)}{(3 + 4 x^4)^2}+4\,.
\ee
This system is described by the potential that is regular on all the real line. 
Its spectrum, up to the  shift equal to four,   is the same as of the harmonic oscillator
$L^{\rm osc}$ except two missing energy levels between $E_1=7$ and $E_4=13$. 
We can modify then the system $L_{(2,3)}$   similarly to the
construction of the half-harmonic oscillator by introducing the infinite
potential barrier at $x=0$ and restricting the domain 
for the semi-infinite interval $(0,+\infty)$. This is equivalent
to removing all the even eigenstates and corresponding eigenvalues
from the spectrum of
the Hamiltonian operator $L_{(2,3)}$.
We denote this intermediate system by $L_{(2,3)}^{1/2}$.
This system has the equidistant 
spectrum with the double distance between energy levels in comparison 
with the spectrum of the quantum harmonic oscillator, 
and its ground state of energy $E_1=7$ is described by the wave function
of the form $\psi=W(2,3,1)/W(2,3)$. Using this ground state
of the obtained system $L_{(2,3)}^{1/2}$
as the seed state, we apply Darboux transformation
to it. The partner system for $L_{(2,3)}^{1/2}$ 
that we produce in this way is exactly  the
system $L_{(1,2,3)}$. 

The generation of the system $L_{(123)}$ 
can also be re-interpreted by taking
$L_1$ as a starting point.
For this we select the state 
$A_1^-\psi_2=\frac{1 + 2 x^2}{x}e^{-x^2/2}$,
singular  at $x=0$,
and  the ground state $A_1^-\psi_3$.
By virtue of iteration properties of the 
DC transformations, one can check
that  with the set of the seed states $\{ A_1^-\psi_2, A_1^-\psi_3, \}$  
we produce the system $L_{(1,2,3)}$.  The 
singular state $A_1^-\psi_2$ can be obtained by 
the same trick we already used 
 by
applying the transformation
$m \rightarrow -(m+1)$  to the 
first excited state of  $L_1$. 
In what follows for us also will 
be important the versions of such states 
obtained by additional 
transformation  $x \rightarrow ix$,
$E\rightarrow -E$.
\vskip0.02cm

It is  
worth to note here that we constructed the 
fourth order ladder operators $\widetilde{\mathcal{B}}^\pm$
by Darboux-dressing the ladder operators of the 
isotonic oscillator $L_2$ though the system $L_{(-3)}$ 
is the isospectral rational deformation of the isotonic oscillator
$L_1$. One can construct the pair of the ladder operators
by Darboux-dressing the ladder operators of the isotonic
oscillator $L_1$, but it turns out to be related 
to the fourth order operators $\widetilde{\mathcal{B}}^\pm$
in a simple way. The analogous phenomena of relation between
different ladder operators obtainable via 
the DC dressing procedure 
 happen also in more complicated cases
of rational deformations of the isotonic oscillator
systems. 
By this reason, let us also consider briefly  the construction
of yet another indicated pair of  ladder operators. 
For this we first write down 
another factorization of the third order intertwining
operator $\A_{(1,2,3)}^-$\,:
$\A_{(1,2,3)}^-=\A^{(1)-}_{(2,3)}A_{(1)}^-$, cf. (\ref{A123A2A}).
Here $A_{(1)}^-=A^-_1$ is the first order differential
operator 
that intertwines $L_0$ and $L_1$, and  
$\A^{(1)-}_{(2,3)}$ is the second order differential 
operator that intertwines the
systems $L_1$ and $L_{(1,2,3)}$.
We can construct now the order six 
ladder operators for the system $L_{(-3)}=L_{(1,2,3)}-8$
by the DC dressing of the ladder operators 
of the isotonic oscillator $L_1$:
$ \hat{\mathcal{B}}^\pm=\A^{(1)-}_{(2,3)}\mathcal{C}^\pm_1\A^{(1)+}_{(2,3)}$.
However, 
they are related to the forth order ladder operators 
$\widetilde{\mathcal{B}}^\pm=-\mathcal{A}^\pm$  in a simple way:
\be
 \hat{\mathcal{B}}^-=\widetilde{\mathcal{B}}^-(L_{(-3)}+1)\,,\qquad
\hat{\mathcal{B}}^+=\widetilde{\mathcal{B}}^+(L_{(-3)}+5)\,.
\ee
The additional factors of the first order  in Hamiltonian annihilate 
the corresponding non-physical eigenstates of $L_{(-3)}$ of eigenvalues $-1$ and $-5$.
This means that the action of these six order ladder operators
on physical states of the system $L_{(-3)}$
is in fact the same as of the ladder 
operators $\widetilde{\mathcal{B}}^\pm=-\mathcal{A}^\pm$
and $\mathcal{B}^\pm$.

Since the system $L_{(-3)}$ is completely isospectral 
to the isotonic oscillator $L_1$ having ladder operators of 
differential order two, there may appear a question
of whether  other ladder operators of differential order two 
do exist here. One can show, however, that the unique 
Hamiltonian operators admitting the existence of  
ladder operators of differential order two are those of the 
harmonic oscillator and the isotonic oscillator systems
as well as their ``anyonic" modifications of the form
(\ref{defiso}) with coefficient $m(m+1)$ changed there 
for $\nu(\nu+1)$ with $\nu>0$.

Consider  briefly a more  complicated case of the system 
$L_{(-3,-7)}$.
This is  an isospectral rational extension
of 
 $L_2^{\rm iso}$ with potential given by Eq. (\ref{V(-3,-7)}).
  The mirror diagram gives  us an equivalent way to obtain the same, up to a shift, system 
 by using the set of the seed states  $(1,2,3,5,6,7)$\,:
$L_{(1,2,3,5,6,7)}=L_{(-3,-7)}+16$, 
where $16=4\Delta E^{\rm iso}$.
The intertwining operators $\A^{-}_{(-3,-7)}$ and 
$\A^{-}_{(1,2,3,5,6,7)}$
satisfy  the relations $
\A^{-}_{(-3,-7)}L_0^{\rm iso}=L_{(-3,-7)}
\A^{-}_{(-3,-7)} 
$ and $
\A^{-}_{(1,2,3,5,6,7)}L_0^{\rm iso}=
L_{(1,2,3,5,6,7)}\A^{-}_{(1,2,3,5,6,7)} 
$.
Since here we also have two  schemes, the procedure developed in the previous example is applicable for the 
construction of the ladder operators of the types  $\mathcal{A}^{\pm}$, $\mathcal{B}^\pm$ and $\mathcal{C}^\pm$. 
Besides, this system can be obtained by taking as a starting point the isotonic 
oscillator $L_4$ to which we apply the DC transformation based on
the seed states 
 $\A_4^-\psi_2$ and  $\A_4^-\psi_6$,  which  are
singular but have  no zeros on the positive half-line. 
So, by dressing  the operators $\mathcal{C}_4^\pm$ with
the intertwining operators corresponding to this transformation, we also can 
assemble a pair of the 
ladder operators of the type 
 $\widetilde{\mathcal{B}}^\pm$.

\vskip0.1cm

The described properties of the two particular examples 
of the rationally extended isotonic oscillator systems 
can be generalized and summarized as follows.
 No matter what set of the $m$ odd non-physical eigenstates of 
 the quantum harmonic oscillator $L^{\rm osc}$ 
 we select,  the lower order ladder operators
 $\mathcal{A}^\pm$ obtained by Darboux-dressing 
 of the ladder operators of the half-harmonic oscillator 
 are spectrum-generating operators 
 for  the rationally deformed isotonic oscillator system.
 They commute for a polynomial of order $2m+1$ 
 in the corresponding  Hamiltonian operator  with which they 
 produce a  deformation of  the conformal $\mathfrak{sl}(2,\R)$ symmetry 
 of the type of $W$-algebra  \cite{BoerHarTji}.
 Other spectrum-generating ladder operators,
 which can be constructed on the basis of other 
 DC schemes via the Darbox-dressing procedure, 
 act on physical states 
 in the same way as the operators $\mathcal{A}^\pm$
 of order $2(m+1)$, and
 are equal to them modulo the multiplicative 
 factor in the form of the polynomial in the  
 Hamiltonian operator of the system.
The ladder operators $\mathcal{C}^{\pm}$
constructed by ``gluing" intertwining operators 
of the two dual  schemes are not spectrum-generating.
Their properties are analogous to those of
the ladder operators $\widetilde{\mathcal{C}}^\pm_m$
in the isotonic oscillator $L^{\rm iso}_m$ with $m>1$.
Particularly, for the isospectral deformation of the 
system $L^{\rm iso}_{l_m+1}$
based on the  set of the 
seed states $(-(2l_1+1),-(2l_2+1),\ldots,
-(2l_m+1))$ with $0\leq l_1<l_2<\ldots<l_m$, $l_m\geq 1$, 
the operator ${\mathcal{C}}^-$
annihilates the lowest $l_m+1$ states in the spectrum of the system.
The Hilbert space of the system $L_{(-(2l_1+1),\ldots,
-(2l_m+1))}$ is separated into $l_m+1$
subspaces in the form of the towers over these states
which are invariant under the  action of the  ladder
operators  ${\mathcal{C}}^\pm$.
However, unlike $\widetilde{\mathcal{C}}^\pm_m$, 
here the ladder operators ${\mathcal{C}}^\pm$  are independent 
from the spectrum-generating 
operators $\mathcal{A}^\pm$.

 \section{Gapped  deformations}\label{SecGapped}

To construct the isotonic oscillator systems and their 
isospectral rational  deformations,
we have introduced infinite potential barrier 
into the quantum harmonic oscillator system 
at $x=0$. In this way we obtained the half-harmonic oscillator $L_0$ and then
we applied to it 
a DC transformation generated on the 
basis of  physical or certain  non-physical  eigenstates
of $L_0$ taken as the seed states. 
This procedure also allowed us to obtain different types
of ladder operators
for corresponding systems.
In order to generate non-isospectral rational deformations  
and  find their ladder operators,  
 we repeat the procedure 
but by changing every time  the half-harmonic oscillator $L_0$ for the half 
of the appropriately chosen  rationally extended quantum harmonic 
oscillator system.
The non-isospectrality corresponds here to the appearance
of $N_g$ ``valence bands",   the highest of which 
is separated by an energy  
gap $n_0\Delta E^{\rm iso}$ from the  
equidistant part of the spectrum with the spacing 
 $\Delta E^{\rm iso}=4$, where $n_0\geq 2$ is integer. 
In the case $N_g\geq 2$, a  separated part of the spectrum 
will contain in addition $N_g-1$ energy gaps  of dimensions  
$n_i\Delta E^{\rm iso}$,
$n_i\geq 2$, $i=1,\ldots,N_g-1$,
which will separate $(N_g-1)$  groups of 
equidistant energy levels   
spaced by $\Delta E^{\rm iso}=4$ inside each valence band with more than one state.
By eliminating then not all but some number of the lowest consecutive 
energy levels from the valence bands by means of a DC transformation,
as a result we obtain a certain deformation of the corresponding 
isotonic oscillator system. 
The general properties of the DC transformations 
will finally allow us  to present the described two-step procedure 
in an equivalent form of a one-step
DC transformation applied to the half-harmonic
oscillator system.

 The described general procedure is realized  explicitly in the following way.
 Consider a quantum system 
  \begin{equation}
 \label{3.1*}
L= -\frac{d^2}{dx^2}+x^2-2(\ln W(l_1,l_1+1,l_2,l_2+1,\ldots,l_m,l_m+1))''\,
\end{equation}  
generated from the quantum harmonic oscillator 
by means of the DC transformation based on a set of the seed states 
$(l_1,l_1+1,l_2,l_2+1,\ldots,l_m,l_m+1)$ 
of    
 $L^{\rm osc}$,
 where we   assume that 
$l_1>0$ and $l_i+1<l_{i+1}$. 
 Quantum system  defined by  (\ref{3.1*}) has an even potential and 
describes a rationally extended quantum harmonic oscillator
system. Its spectrum coincides   with the spectrum 
of the quantum harmonic oscillator except the 
missing energy levels that correspond to the energies of the seed states,
see ref. \cite{CarPly2} for the details.
Then we introduce the  potential barrier into the system (\ref{3.1*}) at $x=0$
by requiring that for $x>0$ the potential of the ``halved" system $L^{1/2}$ 
coincides with the potential of (\ref{3.1*}) while for $x\leq 0$, $V(x)=+\infty$.
This is equivalent to  
demand that the wave functions of the physical states of the 
system $L^{1/2}$   disappear for $x\leq 0$. 
As a result, the eigenstates of the system $L^{1/2}$ 
on the half-line $x>0$
will be described by the odd eigenfunctions of the 
system (\ref{3.1*}).
The spectrum of the system $L^{1/2}$   will be 
given by the numbers $E_{2n+1}=4n+3$,
where $n$ takes all the values $n=0,1,\ldots$,
except those values for which $2n+1$ coincides 
 with any of the  odd integers contained in the 
 set of the numbers which correspond to the indices 
 $(l_1,l_1+1,l_2,l_2+1,\ldots,l_m,l_m+1)$ of the 
 seed states used in the DC transformation 
 for the system   (\ref{3.1*}). 
 This means that  each state with odd index  that appears 
 in the Wronskian 
 in (\ref{3.1*})  
 removes an energy
 level in the spectrum of $L^{1/2}$. As a result, 
by controlling the choice of the set of numbers $l_i$, we can produce a 
halved rationally extended quantum harmonic 
oscillator  system with gaps of arbitrary number and size, at will.  
The even eigenstates are implemented 
in the corresponding DC transformation effectively 
to avoid  singularities in the Hamiltonian operator  (\ref{3.1*}).
After that, we realize an additional DC transformation
based on a set
of consecutive lowest eigenstates of the system $L^{1/2}$
starting from its ground state. These states 
in the set eliminate the corresponding lower lying energy levels 
in spectrum of the partner system  obtained from the system $L^{1/2}$.
If in such a way we eliminate all the energy levels 
from the separated part of the spectrum of $L^{1/2}$,
depending  on the initial choice of the set of numbers $l_i$
in  (\ref{3.1*}),
we obtain just either some isotonic oscillator system
or its some isospectral deformation.
If, however,  we do not eliminate in such a way all the separated lower lying 
energy levels in the spectrum of $L^{1/2}$, we obtain some
non-isospectral, gapped rational extension 
of some isotonic oscillator.  
Using the general properties of the DC 
transformations,
the described procedure with two DC transformations
can be unified just into one DC transformation
based on the corresponding  set of physical eigenstates of the quantum harmonic 
oscillator chosen as seed states. Equivalently, the 
chosen complete set 
of the physical eigenstates of the quantum harmonic oscillator
$L^{\rm osc}$
will correspond to both physical and non-physical eigenstates of the
half-harmonic oscillator $L_0$. A   set of 
physical states of $L_0$ is formed by 
the set of odd eigenstates of $L^{\rm osc}$. 
The set of non-physical eigenstates of $L_0$ 
corresponds here to the even eigenstates of  $L^{\rm osc}$\,; 
they are  formal  eigenstates
of $L_0$ which do not satisfy the boundary condition
for the half-harmonic oscillator  at $x=0$. 

Let us first consider a simple  example taking $m=1$ and fixing 
$l_1=4$. According to (\ref{3.1*}), we obtain  the system
$L_{(4,5)}$ that is a deformed harmonic oscillator with a gap of 
the size $6$ and four separated states given by the corresponding DC map
of the states $\psi_0$, $\psi_1$, $\psi_2$ and $\psi_3$. 
Following the general prescription described above,
 we introduce an infinite barrier at $x=0$
to produce the halved deformed  harmonic oscillator $L_{(4,5)}^{1/2}$
in which all the states with even index are 
removed and   two lowest states are separated from the equidistant part 
of the spectrum by a  gap 
of size $8$. Then we use the ground state of this system,
which is given by the image of the state $\psi_1$,  to generate finally the deformed
isotonic oscillator $L_{(1,4,5)}$ with a potential  
$V_{(1,4,5)}$ given  by Eq. (\ref{V(1,4,5)}).
The system has one separated state 
and  the size  
of the gap is equal to $8$.
This is the minimal possible size  for a gap  in the spectrum 
that can be created by the described general  procedure.
The dual scheme is $(-2,-3,-5) \sim (1,4,5)$, and we can reinterprete 
both these  schemes in terms of the isotonic oscillator. For this 
we generate the system $L_1$ by eliminating 
the ground state $\psi_1$ of the half-harmonic oscillator and then  use
the second excited state of this system given by
$A_{1}^{-}\psi_{5}(x)$ together with the non-physical state
$A_{1}^{-}\psi_{4}(x)$, which is singular at zero.
The use of both indicated states allows us to generate a system to be 
regular on the positive half-line.  In a similar way we also can re-interprete 
the negative scheme by generating the system $L_3$ 
with the help of the auxiliary scheme $(-1,-3,-5)$. Then we 
apply to it additional DC transformation
with the seed states $\A_{-3}^{-}\widetilde{\psi_1^-}=\mathcal{L}_2^{(-7/2)}(x^2) x^{-3}e^{-x^2/2}$
and $\A_{-3}^{-}\psi_2^-=\mathcal{L}_1^{(-7/2)}(-x^2) x^{-3}e^{x^2/2}$,
where $\mathcal{L}_2^{(-7/2)}(x^2)$ has no zeros on the positive half-line $x>0$,
and  $\mathcal{L}_1^{(-7/2)}(-x^2)$  has one zero there.
The first state can be obtained by applying the transformation $m\rightarrow -(m+1)$ with $m=3$
to the second exited state of $L_3$, while the second state is generated  
by  applying  the same transformation to the first excited state
with subsequent change $x\rightarrow ix$.
  As a result  we produce the
same, up  to a displacement of the potential for an additive constant,  final system.
If we look step by step, in this last scheme the image of $\widetilde{\psi_1^-}$
generates an isospectral deformation, under which the image of 
$\psi_2^-$, given by  
\be
\phi(x)=
\frac{W(\A_{-3}^{-}\widetilde{\psi_1^-},\A_{-3}^{-}\psi_2^-)}{\A_{-3}^{-}\widetilde{\psi_1^-}}=
e^{x^2/2}\frac{15 + 10 x^2 - 4 x^4 + 8 x^6}{ x^{2}(15 + 
   4 x^2 (3 + x^2))}\,,
   \ee
    deformes additionally 
   the potential by introducing a new ground state 
     $1/\phi(x)$.

It is worth to note  that we can construct  
another system to be completely  isospectral to
 $L_{(1,4,5)}$ by taking the schemes  
$(1,5,6)\sim (-2,-3,-4,-6)$ (that corresponds to $m=1$ and  $l_1=5$ in Eq.  (\ref{3.1*})), 
whose potential is given by  Eq. 
(\ref{V(1,5,6)}).
As we shall see below,  the
 dual schemes with even highest  even index in the  positive scheme 
have some peculiarity  in the construction of the ladder operators.

\begin{figure}[hbt]
\begin{center}
\includegraphics[scale=0.30]{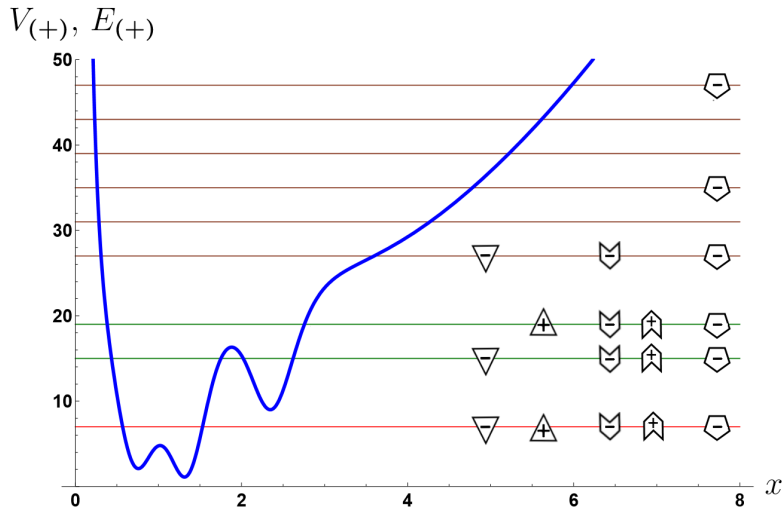}
\caption{Potential (\ref{V(x)long}) of the system (\ref{(1,4,5,10,11)}).
The energy levels of the corresponding physical states
annihilated by ladder operators $\mathcal{B}^-$,
$\mathcal{B}^+$, $\mathcal{A}^-$, $\mathcal{A}^+$, and $\mathcal{C}^-$ are indicated 
from left to right.} 
\label{Figure2}
\end{center} 
\end{figure}

In order to see how the described procedure 
works in general case, 
and to reveal and discuss the peculiarities related to the construction and nature
of the ladder operators, we  consider a more complicated
concrete  nontrivial example. 
Let us take $m=2$ and choose $l_1=4$ and $l_2=10$ in  (\ref{3.1*}).
We then obtain  
the rationally extended quantum harmonic oscillator system 
described by the Hamiltonian operator 
$L_{(4,5,10,11)}=-\frac{d^2}{dx^2}+x^2-2(\ln W(4,5,10,11))''$.
The ground state in the halved system $ L_{(4,5,10,11)}^{1/2}$
corresponds to the DC transformation 
of the eigenstate $\psi_1$ of the quantum harmonic oscillator 
which has  the form $\psi=W(4,5,10,11,1)/W(4,5,10,11)$.
Realizing the Darboux transformation of $ L_{(4,5,10,11)}^{1/2}$
on the basis of this $\psi$ chosen as the seed state,
we obtain the system on the half-line that  equivalently 
can be presented by  the Hamiltonian operator
\begin{equation}\label{(1,4,5,10,11)}
L_{(1,4,5,10,11)}=-\frac{d^2}{dx^2}+x^2-2(\ln W(1,4,5,10,11))''\,.
\end{equation}
This Hamiltonian corresponds to the DC transformation 
of the half-harmonic oscillator $L_0$  generated on the basis 
of the set of its physical and non-physical eigenstates
$(1,4,5,10,11)$. The system (\ref{(1,4,5,10,11)})
represents a non-isospectral rational extension of the 
isotonic oscillator system $L_1$.  The 
explicit analytical form of the potential 
 is given by
Eq. (\ref{V(x)long}).
The graph of this potential and the quantum spectrum of the system
(\ref{(1,4,5,10,11)}) are shown on Figure  \ref{Figure2}.
The potential has three local minima and the system supports
three separated states in its spectrum which are organized in two 
``valence bands"  of one and two states.
By  the method 
of the mirror diagram, we find that the DC scheme (-2,-3,-4,-5,-8,-9,-11) 
produces the same Hamiltonian operator but shifted by  a constant,
$L_{(+)}-L_{(-)}=6\Delta E^{\rm iso}=24$, see Fig. \ref{Figure1}. 
Here, for simplicity,  we introduced the notations
 $L_{(+)}=L_{(1,4,5,10,11)}$ and $L_{(-)}=L_{(-2,-3,-4,-5,-8,-9,-11)}$. 
The fact that the mutual shift  of both Hamiltonians is proportional to the 
difference of two consecutive energy levels  in the spectrum  of the isotonic oscillator 
allows us to use below exactly the same  rule  for
the construction of the ladder
 operators of the type $\mathcal{C}^{\pm}$
 as in the previous section.  
As we shall see,  the number of physical  
states annihilated by the lowering operator
$\mathcal{C}^-$ in this case is equal exactly to six.    
Later, we also shall see that in some cases  
of the  rational gapped deformations of the isotonic oscillator systems, 
the mutual shift 
of the corresponding Hamiltonian operators 
can be equal to the half-integer 
multiple of $\Delta E^{\rm iso}$, and then 
the procedure for the 
construction of the ladder operators of the 
type $\mathcal{C}^{\pm}$
will require some modification.

 In the DC construction  of the Hamiltonian operator $L_{(+)}$, 
 the energy levels corresponding 
 to the physical seed eigenstates of the half-harmonic oscillator $L_0$
 were removed from the spectrum  producing two gaps.
 In the equivalent  system $L_{(-)}$ based on non-physical seed eigenstates
 of $L_0$, the  energy levels were added under the lowest energy 
of the ground  state of $L_0$. 
 The fifth order differential operators that intertwine the half-harmonic oscillator
 $L_0$ with  $L_{(+)}$ are
 \be
\A^-_{(+)}=\A_{(1,4,5,10,11)}^{-}\,, \qquad  
\A^+_{(+)}=\A_{(1,4,5,10,11)}^{+}\,. 
\ee
The seventh order differential operators 
intertwining  $L_0$ with $L_{(-)}$ are
\be
\A_{(-)}^{-}=\A_{(-2,-3,-4,-5,-8,-9,-11)}^-\,,  \qquad
\A_{(-)}^{+}=
\A_{(-2,-3,-4,-5,-8,-9,-11)}^{+}\,. 
\ee
The three lowest physical states of the system (\ref{(1,4,5,10,11)})
which correspond to the three separated energy levels 
can be presented in two equivalent forms
$ \A_{(-)}^{-}\widetilde{\psi_8^-}=\A^-_{(+)}\psi_3$,
 $ \A_{(-)}^{-}\widetilde{\psi_4^-}=\A^-_{(+)}\psi_7$, 
 $\A_{(-)}^{-}\widetilde{\psi_2^-}=\A^-_{(+)}\psi_9$, 
 where equalities are modulo a nonzero constant multiplier.
We have  here the intertwining relations
\be
\A^-_{(+)}L_0=L_{(+)}\A^-_{(+)}=(L_{(-)}+24)\A^-_{(+)}\,, \qquad 
\A_{(-)}^{-}L_0=L_{(-)}\A_{(-)}^{-}=(L_{(+)}-24)\A_{(-)}^{-}, 
\ee
and the  conjugate relations for $\A^+_{(+)}$ and $\A_{(-)}^{+}$.

From the point of view of the isotonic oscillator we can  re-interprete 
the positive scheme starting from $L_1$. 
One can check that the state
$A_1^-\psi_4=x^{-1}\mathcal{L}_2^{(-3/2)}(x^2)e^{-x^2/2}$ 
is obtainable by applying the transformation   $m \rightarrow -m-1$ with $m=1$
to the second excited state  while   $A_{1}^-\psi_{10}=x^{-1}{\mathcal L}_5^{(-3/2)}(x^2)e^{-x^2/2}$ 
is generated by the same transformation from the fifth excited state.
By iterative properties  of the DC transformations we can see 
that the selection of the states
 $\{A_1^-\psi_4, A_1^-\psi_{10}, A_{1}^-\psi_5, A_1^-\psi_{11}\}$ 
 produces our deformed system. 
The negative scheme here 
 also can be re-interpreted  by constructing first the isotonic system  $L_6$
 by using  the set of six odd non-physical states  $\{-1,-3,-5,-7,-9,-11\}$ 
 and then taking $\A_{-6}^{-}\widetilde{\psi_1^-}$, $\A_{-6}^{-}\widetilde{\psi_7^-}$, 
 $\A_{-6}^{-}{\psi_2^-}$, $\A_{-6}^{-}{\psi_4^-}$ and  
 $\A_{-6}^{-}{\psi_8^-}$
 as the seed states. 
 {}From the point of view of 
$L_6$, these states can be obtained from some physical states by applying the symmetry
transformations of the isotonic oscillator as in the previous considered cases, 
and if we trace step by step, we see that  
the states $\A_{-6}^{-}\widetilde{\psi_1^-}$  and  $\A_{-6}^{-}\widetilde{\psi_7^-}$ produce some isospectral deformation of 
 $L_4$. In this new intermediate system three other states introduce three separated states below
 the equidistant part of the spectrum by producing finally 
 our deformed oscillator system (\ref{(1,4,5,10,11)}).

Let us turn now to the construction of the ladder
operators for the system under consideration.
As in the previous section, we  begin with the  
Darboux-dressed ladders operators,
and  
first consider 
 the operators of the type $\widetilde{\mathcal{B}}^{\pm}$.
 They  are  constructed from 
the ladder operators $\mathcal{C}_1^{\pm}$  of the isotonic oscillator
 $L_1$, and  are the lowest  (tenth) order ladder 
 operators we can obtain by
  using the method of Darboux-dressing\,: 
  \begin{equation}\label{tildeA(4,5,10,11)}
\widetilde{\mathcal{B}}^{\pm}=\A_{(4,5,10,11)}^{(1)-}\mathcal{C}_1^{\pm}\A_{(4,5,10,11)}^{(1)+}\,,\qquad
[L_{(+)},\widetilde{\mathcal{B}}^{\pm}]=\pm\Delta E^{\rm iso}\widetilde{\mathcal{B}}^{\pm}\,.
\end{equation}
The fourth order differential operators
 $\A_{(4,5,10,11)}^{(1)\pm}$ 
   intertwine  
   the Hamiltonian
 operators  $L_1$ and $L_{(+)}$
 and realize the mapping between their corresponding eigenstates. 
 The numbers in  the subscript  
 indicate the 
 eigenstates 
 of the harmonic oscillator used as 
the  seed states of the DC transformation, 
and the superscript
marks the system to 
which those states are mapped. 
In correspondence with this notation and properties
of the DC transformations, we have here 
the  
 equality
$\A^{(1)-}_{(4,5,10,11)}(A_{(1)}^-\psi)=\A^-_{(1,4,5,10,11)}\psi$,
where $\psi$ is an arbitrary eigenstate of the half-harmonic 
oscillator Hamiltonian $L_0$.
The kernel  of  $\widetilde{\mathcal{B}}^{-}$ can be identified by calculating the product
 $\widetilde{\mathcal{B}}^{+}\widetilde{\mathcal{B}}^{-}$, while  the kernel of
 $\widetilde{\mathcal{B}}^{+}$ is found from the product 
$\widetilde{\mathcal{B}}^{-}\tilde{\mathcal{B}}^{+}$.   
We find 
\begin{eqnarray}
\ker\widetilde{\mathcal{B}}^{-}=\text{span}\,\{
\A_{(-)}^{-}\widetilde{\psi_{11}^{-}},\,
\A_{(-)}^{-}\widetilde{\psi_8^{-}},\,
\A_{(-)}^{-}\psi_{7}^{-},\,
\A_{(-)}^{-}\psi_6^{-},\,
\A_{(-)}^{-}\widetilde{\psi_5^{-}},\,\nonumber\\
 \A_{(-)}^{-}\widetilde{\psi_4^{-}},\,
\A_{(-)}^{-}\psi_1^{-},\,
\A_{(-)}^{-}\psi_0^{-},\,
\A_{(-)}^{-}\psi_0,\, 
\A_{(-)}^{-}\psi_1\,
\}. 
\end{eqnarray}
The kernel of $\widetilde{\mathcal{B}}^{-}$ contains three physical states  
$\A_{(-)}^{-}\widetilde{\psi_8^{-}}$,  $\A_{(-)}^{-}\widetilde{\psi_4^{-}}$ and 
$\A_{(-)}^{-}\psi_1=\A_{(+)}^{-}\psi_{13}$.  
These are the  ground state, the lower state in the two-states  ``valence band"
(that is also the first excited state of the system), 
 and the lowest state in the equidistant part of the spectrum, see Figure \ref{Figure2}. 
 They are eigenstates of  $L_{(-)}$ of energies $-17$ and $-9$
 and $3$, respectively.
The kernel 
\begin{eqnarray}
\ker\widetilde{\mathcal{B}}^{+}=\text{span}\,\{
\A_{(-)}^{-}\psi_{13}^{-},\,
\A_{(-)}^{-}\psi_{10}^{-},\,
\A_{(-)}^{-}\widetilde{\psi_9^{-}},\,
\A_{(-)}^{-}\widetilde{\psi_8^{-}},\,  
\A_{(-)}^{-}\psi_7^{-},\, \nonumber\\
\A_{(-)}^{-}\psi_6^{-},\,
\A_{(-)}^{-}\widetilde{\psi_3^{-}},\,
\A_{(-)}^{-}\widetilde{\psi_{2}^{-}},\,
\A_{(-)}^{-}\psi_1^{-},\, 
\A_{(-)}^{-}\psi_{0}^{-}
\}\,,
\end{eqnarray}
contains two physical states
$\A_{(-)}^{-}\widetilde{\psi_8^{-}}$ and
 $\A_{(-)}^{-}\widetilde{\psi_2^{-}}$, which are identified as the 
 ground state and the second excited state of the system. 
 Both states are the upper states in the 
 one-state and the two-states valence bands. This is correlated 
with the property that  
the jump produced by  the raising operator $\widetilde{\mathcal{B}}^{+}$ 
 is not sufficient  to bridge the  corresponding gaps
in the spectrum. The operators $\widetilde{\mathcal{B}}^{\pm}$ 
cannot connect the part of the spectrum 
corresponding to the separated states 
with the states in the equidistant part of the spectrum.
Also, they do not connect  the states from different valence 
bands. 
These ladder operators  do not  form the set of  the spectrum-generating  operators
since they do not allow us to generate arbitrary state in the spectrum from
any other arbitrary chosen fixed physical eigenstate. 
All the space of the states of the system is separated into three 
bands,  
in each of which 
the operators $\tilde{\mathcal{B}}^{\pm}$ act 
irreducibly. These ladder operators commute for certain 
polynomial of order nine in the Hamiltonian $L_{(-)}$.

Like in the isospectral case considered in the previous section, 
here we have two ways to realize 
Darboux-dressing of the ladder operators 
 $-\mathcal{C}^\pm_0=(a^{\pm})^{2}$ of the half-harmonic oscillator system
 $L_0$.
Using for this purpose the intertwining operators $\A^\pm_{(+)}$,
 we obtain differential operators of order twelve:
\begin{equation}\label{BcalAladder}
\mathcal{B}^{\pm}=\A^-_{(+)}(a^{\pm})^{2}\A^+_{(+)}\,,\qquad [L_{(-)},\mathcal{B}^{\pm}]=\pm\Delta E^{\rm iso}\mathcal{B}^{\pm}\,.
\end{equation}
With the help of the  product 
$\mathcal{B}^{+}\mathcal{B}^{-}=(L_{(-)}+21)^{2}
\widetilde{\mathcal{B}}^{+}\widetilde{\mathcal{B}}^{-}$, 
we find  the kernel of the operator $\mathcal{B}^{-}$\,:
$\ker \mathcal{B}^{-}=\text{span}\,
\{ \A_{(-)}^{-}\psi_{10}^{-},\,\A_{(-)}^{-}
\widetilde{\psi_{10}^{-}},\ker \widetilde{\mathcal{B}}^{-}\}$. 
The first two states in kernel of $\mathcal{B}^{-}$ 
are non-physical eigenstates of the operator $L_{(-)}$ of energy
$E_{(-)}=-21$.  As a result we see that the ladder operator 
$ \mathcal{B}^{-}$ 
annihilates exactly the same three physical eigenstates 
as the ladder operator $\widetilde{\mathcal{B}}^{-}$,
and the other nine eigenstates from the  kernel  of $\mathcal{B}^{-}$ 
are non-physical.
{}From  here 
one can suspect that  the operators $\mathcal{B}^{\pm}$
and $\widetilde{\mathcal{B}}^{\pm}$ are related
in a simple way, and we find 
\be
\mathcal{B}^{-}=\widetilde{\mathcal{B}}^{-}
(L_{(-)}+21)\,,\qquad
 \mathcal{B}^{+}=(L_{(-)}+21)\widetilde{\mathcal{B}}^{+}\,. 
 \ee
The kernel of 
$\mathcal{B}^{+}$ is spanned 
by the kernel of $\widetilde{\mathcal{B}}^{+}$ and 
two non-physical eigenstates 
$\A_{(-)}^{-}\psi_{12}^{-}$ and $\A_{(-)}^{-}\widetilde{\psi_{12}^{-}}$
of $L_{(-)}$ of eigenvalue
$E_{(-)}=-25$.

We also can construct the  
ladder operators by
using the intertwining operators $\A_{(-)}^{\pm}$,
\begin{equation}\label{AaADarboux}
 \mathcal{A}^{\pm}=\A_{(-)}^{-}(a^{\pm})^{2}\A_{(-)}^{+}\,,
\qquad  
[L_{(+)},\mathcal{A}^{\pm}]=\pm\Delta E^{\rm iso}\mathcal{A}^{\pm\,}\,.
\end{equation}
These  are also not spectrum-generating operartors
because the leap they make does not allow  to overcome the gaps.  
On the other hand, since these are differential operators of order sixteen, 
by analogy with the case of rationally extended quantum harmonic oscillator
systems \cite{CarPly2},
one could expect  that  they annihilate more
physical states in comparison with the ladder operators
 $\mathcal{B}^{\pm}$ and $\widetilde{\mathcal{B}}^{\pm}$. 
To identify the kernels of the operators $ \mathcal{A}^{\pm}$,
it is more convenient to work  with 
the operator $L_{(+)}$ instead of $L_{(-)}$.
By calculating the alternative products of 
 of the operators $ \mathcal{A}^{\pm}$, one can see  that 
they can be written in terms of $\widetilde{\mathcal{B}}^\pm$: 
\be
\mathcal{A}^{-} =\widetilde{\mathcal{B}}^{-}(L_{(+)}-5)
(L_{(+)}-17)(L_{(+)}-19)\,,\qquad
 \mathcal{A}^{+}=(\mathcal{A}^{-})^\dagger\,.
\ee 
{}From here one finds that the raising operator $\mathcal{A}^{+}$ detects all the 
 states in both  separated valence bands by annihilating them. In addition
 to the indicated physical states,
 the lowering operator $\mathcal{A}^{-}$ also annihilates  the lowest state in the 
 half-infinite equidistant part of the spectrum.

Therefore, the essential difference of the non-isospectral rational deformations
of the isotonic oscillator systems from their  isospectral rational extensions is that
there is no pair of spectrum-generating ladder operators constructed  via the Darboux-dressing procedure. 
This situation is similar to that in the rationally extended quantum harmonic oscillator
systems \cite{CarPly2}.

We now construct 
 the ladder operators $\mathcal{C}^\pm$ 
  by ``gluing" 
 the intertwining operators of different types. As in the case of the isospectral deformations,
 they also will not be the spectrum-generating operators, but
 together with any pair of the ladder operators 
 $\widetilde{\mathcal{B}}^{\pm}$, ${\mathcal{B}^{\pm}}$, or $\mathcal{A}^\pm$
 they will form the set of the spectrum-generating operators.
So, let us consider the differential  operators  of order twelve,  
\begin{equation}\label{C+-AAdefin}
\mathcal{C}^{-}=\A_{(-)}^-\A_{(+)}^{+} \,,
\qquad 
\mathcal{C}^{+}=\A_{(+)}^{-}\A_{(-)}^+\,,
\qquad
[L_{(-)},\mathcal{C}^{\pm}]=\pm 6\Delta E^{\rm iso}\mathcal{C}^{\pm}\,.
\end{equation}
They are independent
from the ladder operators constructed via the Darboux-dressing procedure,
and their commutator $[\mathcal{C}^-, \mathcal{C}^+]$ is a 
certain polynomal of order $11$ in the Hamiltonian $L_{(-)}$.
The  operators $\mathcal{C}^{\pm}$
 divide the space of the  states of the system 
into  six infinite subsets on which they act irreducibly. 
The same number six  here also corresponds to the number of physical 
states 
$\A_{(-)}^-\widetilde{\psi_8^{-}}$,
$\A_{(-)}^-\widetilde{\psi_4^{-}}$, 
$\A_{(-)}^-\widetilde{\psi_2^{-}}$, 
$\A_{(-)}^-\psi_{1}$,
$\A_{(-)}^-\psi_{5}$,
$\A_{(-)}^-\psi_{11}$
annihilated by the lowering  operator $\mathcal{C}^{-}$.
The $\mathcal{C}^{-}$ transforms a physical 
eigenstate into a physical eigenstate
by making it skip six levels below.
If the arrival level is not in the spectrum, then
 that state is  annihilated
by $\mathcal{C}^{-}$.
 
\begin{figure}[H]
\begin{center}
\includegraphics[scale=0.55]{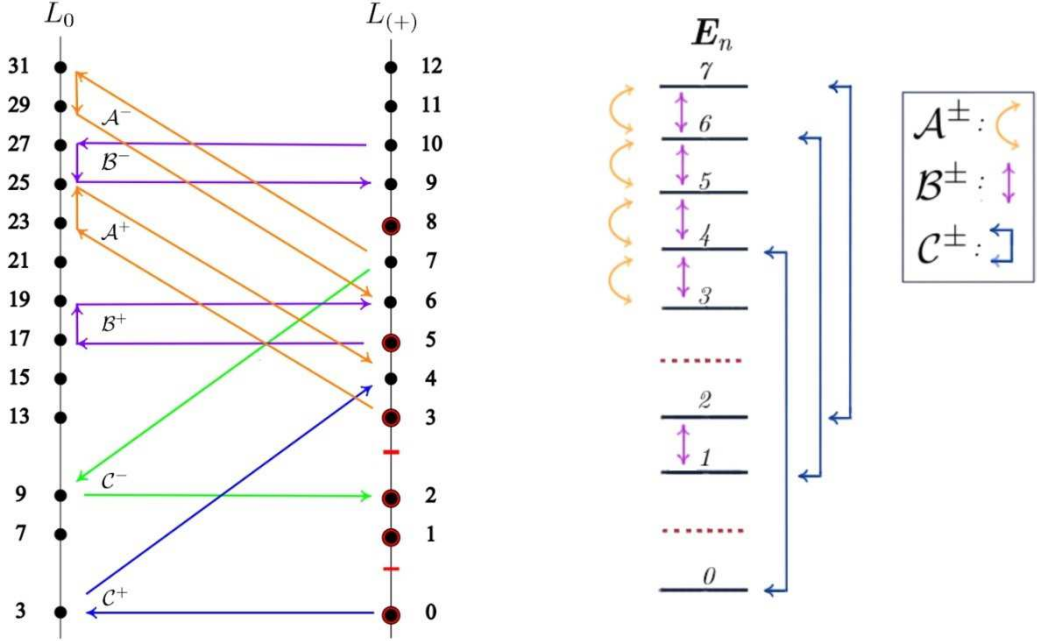}
\end{center} 
\caption{{\label{Figure3}}
Left panel: The numbers on the left correspond to 
the indices of the physical eigenstates $\psi_{2l+1}$ of the half-harmonic oscillator 
that are mapped ``horizontally" by  operator $\A^-_{(+)}$ into  
eigenstates $\Psi_n$ of the system (\ref{(1,4,5,10,11)}). 
Lines show the action 
of the ladder operators coherently with their structure 
(\ref{AaADarboux}), (\ref{BcalAladder}) and 
(\ref{C+-AAdefin}). The marked set of the states $0,\, 1,\, 2,\, 3,\, 5,\,8 $ 
on the right  corresponds to six eigenstates of $L_{(+)}$ annihilated by $\mathcal{C}^-$.
Right panel: Horizontal lines correspond to the  energy levels of $L_{(+)}$. 
Upward and downward arrows  represent the action of the rising and lowering ladder operators, 
respectively. 
Following appropriate  paths, any eigenstate can be transformed into any other eigenstate
by applying subsequently the corresponding ladder operators.}  
\end{figure}

The  operator $\mathcal{C}^{+}$ does not annihilate any physical state here.
Like $\mathcal{C}^{-}$, it connects the separated states with the equidistant 
part of the spectrum.
As a result, the pair of  the ladder operators $\mathcal{C}^{\pm}$ 
together with any pair of the ladder operators $\widetilde{\mathcal{B}}^\pm$, 
$\mathcal{B}^\pm$ or $\mathcal{A}^\pm$ allows us to transform any physical
state into any physical state. For instance, 
the separated ground state  $\A_{(-)}^-\widetilde{\psi_8^{-}}$ of the system
 can be transformed 
into the second  state  $\A_{(-)}^-\widetilde{\psi_2^{-}}$
in the second valence band   by application
of the product of the ladder operators 
$\mathcal{C}^-(\widetilde{\mathcal{B}}^+)^3\mathcal{C}^+$.
Then, the ground state can be  transformed into 
the first excited state $\A_{(-)}^-\widetilde{\psi_4^{-}}$, 
 being the lowest state in the 
two-states valence band, by the operator $\widetilde{\mathcal{B}}^-
\mathcal{C}^-(\widetilde{\mathcal{B}}^+)^3\mathcal{C}^+$.
The ground state can also  be transformed into the lowest 
state of the two-states valence band by
the operator $\mathcal{C}^-(\mathcal{A}^+)^2\mathcal{C}^+$.
The  application of the  operator $\mathcal{B}^-\mathcal{C}^+$	
to  the ground state transforms it into the lowest state 
$\A_{(-)}^-\psi_{1}$ of the equidistant part of the spectrum.
So, the indicated pairs of the ladder operators form the 
sets of the spectrum-generating operators of the system.
Figure \ref{Figure3} illustrates the action of the ladder operators.
Now, consider briefly the systems that can be obtained from 
the system we have just described by taking  its separated states 
as the seed states to generate the corresponding new
DC  transformations.
Each time, the incorporation of a separated state
into a set of the seed states deforms the potential and 
decreases in one the number of its local minima.
As a result, this reduces in one the number of separated
states in the spectrum of the corresponding system.

If in the set of the seed states for the DC transformation 
we include additionally the first excited state 
of the halved rationally extended quantum  harmonic 
oscillator system $L^{1/2}_{(4,5,10,11)}$,
or, in an equivalent way,  if we realize a Darboux transformation of the system 
$L_{(1,4,5,10,11)}$ by using its lowest physical state 
$\A^-_{(+)}\psi_3 =  \A_{(-)}^{-}\widetilde{\psi_8^-}$
from
 the one-state valence band,
we obtain the system  
$L_{(1,3,4,5,10,11)}$. 
It corresponds to a non-isospectral deformation of the isotonic oscillator $L_2$
with the spectrum containing 
two separated states organized into one two-state valence band. 
A dual scheme  is $ (-2,-3,-4,-5,-9,-11)\sim (1,3,4,5,10,11)$.
As before, we simplify notations by denoting
$L_{(+)}=L_{(1,3,4,5,10,11)}$ and 
$L_{(-)}=L_{(-2,-3,-4,-5,-9,-11)}$,
for which, again,   $L_{(+)}-L_{(-)}=24$. 
The sixth order differential 
 operators that intertwine 
$L_{(+)}$ and $L_{(-)}$ with $L_0$ are 
 $\A_{(+)}^\pm=\A_{(1,3,4,5,10,11)}^\pm$ and 
$\A_{(-)}^\pm=\A_{(-2,-3,-4,-5,-9,-11)}^\pm$.
 With their help we construct 
the ladder operators  $\mathcal{B}^\pm$
and $\mathcal{A}^\pm$ by the DC dressing of 
the ladder operators $-\mathcal{C}^\pm_0=(a^\pm)^2$ 
of $L_0$ as it was done for the system (\ref{(1,4,5,10,11)}).
The 
order $12$ ladder operators $\mathcal{C}^\pm$
are also constructed in the form 
(\ref{C+-AAdefin}) using the sixth order 
intertwining  operators  $\A_{(+)}^\pm$ and $\A_{(-)}^\pm$.
The analogs  of the ladder operators 
(\ref{tildeA(4,5,10,11)}) are constructed 
here by the DC dressing 
of the ladder operators of the isotonic oscillator
$L_2$, 
$\widetilde{\mathcal{B}}^{\pm}=\A_{(4,5,10,11)}^{(1,3)-}
\mathcal{C}_2^{\pm}\A_{(4,5,10,11)}^{(1,3)+}$.
The $\A_{(4,5,10,11)}^{(1,3)-}$ and
$\A_{(4,5,10,11)}^{(1,3)+}=(\A_{(4,5,10,11)}^{(1,3)-})^\dagger$
are the fourth order operators  intertwining  the system $L_{(1,3,4,5,10,11)}$
with the isotonic oscillator $L_2$. They   
 are given by relations of the form  
 \be
 \A_{(4,5,10,11)}^{(1,3)-}(\A_{(1,3)}^-\psi)=
 \A^-_{(\A_{2}^-4,\A_{2}^-5,\A_{2}^-10,\A_{2}^-11)}(\A_{2}^-\psi)
=\A^-_{(1,3,4,5,10,11)}\psi\,,
\ee
 where 
$\A_{(1,3)}^-=\A_2^-$ is the operator  (\ref{AAm}) with $m=2$ 
that  intertwines $L_0$ and $L_2$.
The ladder operators ${\mathcal{B}}^{\pm}$
and $\widetilde{\mathcal{B}}^{\pm}$ act on physical states in the same way.
The lowering operators annihilate the ground state 
being the lower state in the two-states valence band and  
the lowest  
state in the equidistant part of the spectrum.
The raising operators annihilate the first excited 
state that corresponds
to the upper energy level in the valence band.
The raising  operator ${\mathcal{A}}^{+}$
annihilates both states in the valence band of energies $E_{(+)}=15$
 and $E_{(+)}=19$
whereas the lowering operator ${\mathcal{A}}^{-}$
annihilates in addition the lowest state  of energy $E_{(+)}=27$ in the equidistant 
part of the spectrum.
One can  find the  relations between these ladder operators
as it was done for the system (\ref{(1,4,5,10,11)}).
The lowering operator $\mathcal{C}^-$ annihilates both states
 $\A_{(+)}^{-}\psi_7$ and $\A_{(+)}^{-}\psi_9$
of the valence band,  the lowest state  $\A_{(+)}^{-}\psi_{13}$ in the equidistant part
of the spectrum, 
and three more physical states  there which are
 $\A_{(+)}^{-}\psi_{15},\, \A_{(+)}^{-}\psi_{17}$, and  $\A_{(+)}^{-}\psi_{23} $
 of energies  $E_{(+)}=31$, $E_{(+)}=35$ and $E_{(+)}=47$. Besides, it 
 also annihilates  six  non-physical eigenstates.
The kernel of the raising operator $\mathcal{C}^+$ 
contains only non-physical states.
As in the previous system $L_{(1,4,5,10,11)}$,
the ladder operators $\mathcal{C}^\pm$ together
with any of the pairs of ladder operators $\widetilde{\mathcal{B}}^{\pm}$, 
${\mathcal{B}}^{\pm}$, or ${\mathcal{A}}^{\pm}$
form the sets of the spectrum-generating operators
of  the system $L_{(1,3,4,5,10,11)}$.

One can continue in the same vein and obtain the system 
$L_{(1,3,4,5,7,10,11)}=L_{(-2,-3,-5,-9,-11)}+24$ that has the same spectrum 
as the system $L_{(1,4,5,10,11)}$ but with the omitted two lowest energy levels.
This is a non-isospectral rational deformation of the
isotonic oscillator $L_3$ with one energy level 
separated by a distance $2\Delta E^{\rm iso}=8$ from the
equidistant part of the spectrum.
Finally, we  obtain the system $L_{(1,3,4,5,7,9,10,11)}=L_{(-3,-5,-9,-11)}+24$
by eliminating the unique separated energy level 
from the spectrum of the system $L_{(1,3,4,5,7,9,10,11)}$. This is 
an  isospectral deformation of the isotonic oscillator $L_4$.
\vskip0.2cm

All the described picture is generalized directly in the case when
the index of the last seed state used in the corresponding
DC transformation is odd. 
Then the corresponding scheme based on physical eigenstates 
of $L_0$ is of the form
$(\ldots,2l_m,2l_m+1)$, and the dual scheme
is  $(\ldots,-(2l_m+1))$.
Following the same notation as we used 
in the particular examples, the Hamiltonian operators generated in these  two 
dual schemes are shifted by the distance equal to the separation $\Delta E^{\rm iso}=4$
of energy levels in the equidistant part of the spectrum times
integer number $l_m+1$\,: 
$L_{(+)}-L_{(-)}=4l_m+4$. 
The operators $\A^\pm_{(+)}$ that intertwine the Hamiltonian $L_{(+)}$
with $L_0$,
\be
\A^-_{(+)}L_0=L_{(+)}\A^-_{(+)}=(L_{(-)}+4l_m+4)\A^-_{(+)}\,,
\ee
are of differential order $n_+$, where $n_+$ corresponds to the number
of  seed states in the ``positive"  scheme $(\ldots,2l_m,2l_m+1)$.
The operators $\A^\pm_{(-)}$ intertwining $L_{(-)}$ with $L_0$,
\be
\A^-_{(-)}L_0=L_{(-)}\A^-_{(-)}=(L_{(+)}-4l_m-4)\A^-_{(-)}\,,
\ee
 in this case 
are of differential order $2l_m+2-n_+$. The ladder operators
$\mathcal{A}^\pm=\A^-_{(-)}(a^\pm)^2\A^+_{(-)}$ 
have differential order $2(2l_m+2-n_+)+2$. The 
ladder operators $\mathcal{B}^\pm$  constructed 
by Darboux-dressing of ladder operators of the half-harmonic 
oscillator $L_0$ by means of intertwining operators $\A^\pm_{(+)}$
are of differential order $2n_++2$.
These ladder operators satisfy commutation relations of the
form  (\ref{AaADarboux}) and (\ref{BcalAladder}).
The lowering operator $\mathcal{B}^-$ annihilates all the lowest 
states in each valence band and the lowest state in the 
equidistant part of the spectrum. 
The raising operator $\mathcal{B}^+$ annihilates all the highest 
states in each valence band. 
The raising operator $\mathcal{A}^+$ detects all the states 
in all the valence bands by annihilating them. 
The lowering ladder operator 
$\mathcal{A}^-$ annihilates in addition to all
the separated states also the lowest state in the equidistant
part of the spectrum. 
The ladder operators $\mathcal{C}^\pm$ of the form (\ref{C+-AAdefin}) 
constructed
by gluing intertwining operators from dual schemes 
have a differential order
$2l_m+2$, and commute with Hamiltonian operators of the
system
as follows\,: 
\begin{equation}\label{LCpm(lm+1)}
[L_{(-)}, \mathcal{C}^\pm]=\pm (l_m+1)\Delta E^{\rm iso}\mathcal{C}^\pm\,.
\end{equation}
Like the ladder operator $\mathcal{A}^-$,
the  lowering operator  $\mathcal{C}^-$ annihilates all the 
states in all the valence bands and the lowest state 
in the equidistant part of the spectrum.
In addition, it annihilates certain number of excited states 
in the equidistant part of the spectrum.
The total number of the physical states annihilated
by $\mathcal{C}^-$ is equal to $l_m+1$ that
coincides  with the corresponding coefficient 
in  
 (\ref{LCpm(lm+1)}).
The kernel of the raising  operator $\mathcal{C}^+$
contains no physical states.
The ladder operators $\mathcal{C}^\pm$ together with 
$\mathcal{A}^\pm$ or 
$\mathcal{B}^\pm$ form the set of  spectrum-generating operators.
In these sets of  spectrum-generating operators,
the ladder operators $\mathcal{B}^\pm$ can be substituted 
by the ladder operators $\widetilde{\mathcal{B}}^\pm$.
The operators $\widetilde{\mathcal{B}}^\pm$ are constructed
by  Darboux-dressing of the ladder operators $\mathcal{C}_m^\pm$
in the way how we constructed such operators 
in the considered examples. Here we assume 
that the corresponding Hamiltonians $L_{(+)}$ and $L_{(-)}$
describe a certain rational deformation of the isotonic oscillator
system $L_m$. It may or may not happen that the
operators of the type $\widetilde{\mathcal{B}}^\pm$ of lower differential 
order can be obtained by Darboux-dressing of ladder operators of 
some another  isotonic oscillator system. 
This, however, depends on the concrete rational deformation
of the isotonic oscillator  we have and requires 
a concrete  investigation. 
With respect to the physical states, the ladder operators
$\widetilde{\mathcal{B}}^\pm$ will have the same properties
as the ladder operators ${\mathcal{B}}^\pm$.
\vskip0.1cm

When we have the schemes $(\ldots,2l_m-1,2l_m)\sim (\dots,-2l_m)$
generating a  gapped rational extension 
of some isotonic
oscillator system, the corresponding Hamiltonian operators
associated with them are shifted mutually for the distance 
$L_{(+)}-L_{(-)}=4l_m+2=(l_m+\frac{1}{2})\Delta E^{\rm iso}$, 
that is equal to the half-integer multiple of the energy spacing 
in the equidistant part of the spectrum and
in the valence bands with more than one state.
In this case the procedure related to the construction 
of the ladder operators $\mathcal{A}^\pm$, 
$\mathcal{B}^\pm$  and $\widetilde{\mathcal{B}}^\pm$ 
and their properties are similar to those 
in the systems generated by the schemes 
$(\ldots,2l_m,2l_m+1)\sim (\dots,-(2l_m+1))$.
However, the situation with the construction of
the ladder operators  of the type $\mathcal{C}^\pm$
in this case is essentially different.
We still can construct the operators $\mathcal{C}^\pm$
of the form (\ref{C+-AAdefin}).
Such operators will be of odd differential order 
$2l_m+1$, and their commutation relations  with any of the
 Hamiltonian operators $L_{(+)}$ and  $L_{(-)}$ 
 will be  of the form $[L, \mathcal{C}^\pm]=\pm (4l_m+2)\mathcal{C}^\pm$.
 This means that these operators acting on physical 
 eigenstates of $L$  will produce non-physical eigenstates
 excepting the case when the lowering 
 operator $\mathcal{C}^-$ acts on the states 
 from its kernel.  
  The square of these operators 
 will not have the indicated deficiency and will form together 
 with the ladder operators $\mathcal{A}^\pm$, $\mathcal{B}^\pm$, or
 $\widetilde{\mathcal{B}}^\pm$  the set of the
 spectrum-generating 
 operators. This picture can be compared with 
  the case of the half-harmonic oscillator $L_0$, where 
  the first order differential operators 
$a^\pm$ will have the properties similar to those of the 
described operators $\mathcal{C}^\pm$. 
In this case we can however modify slightly the construction
of the ladder operators of the $\mathcal{C}^\pm$ type 
by taking 
\be\label{Cpmnew}
\widetilde{\mathcal{C}}^-=A_{(-)}^-(a^-)A_{(+)}^+\,,\qquad
\widetilde{\mathcal{C}}^+=A_{(+)}^-(a^+)A_{(-)}^+\,.
\ee 
These ladder operators satisfy the commutation
relations 
$[L_{(\pm)},\widetilde{\mathcal{C}}^\pm]=4(l_m+1)\widetilde{\mathcal{C}}^\pm$,
and transform physical states into physical states.

Note that the 
total number of physical eigenstates from  the 
half-infinite equidistant part of the spectrum which 
belong to  the kernel 
of the ladder operator $\mathcal{C}^-$ 
(or $\widetilde{\mathcal{C}}^-$) in the case 
of the system of the type 
$(\ldots,2l_m+1)$ ( or $(\ldots,2l_m)$)
can be expressed in terms of 
the sizes $n_+$ and $n_-$ 
of the corresponding ``positive" and ``negative" dual schemes
and the total number $n_v$ of the states
in all the valence bands of the deformed gapped 
isotonic oscillator system.
If we denote such a number by $n_\infty$,
we obtain $n_\infty=\frac{1}{2}(n_++n_-)-n_v$
for the system of the type $(\ldots,2l_m+1)$, for which
$2l_m+1=n_++n_--1$, see Section 
\ref{SectionDual}. 
For the system 
with ``positive" scheme of the  type $(\ldots,2l_m)$,
we have $n_\infty=\frac{1}{2}(n_++n_--1)-n_v$.
\vskip0.1cm 

To conclude  this section, let us summarize
the structure of the nonlinearly deformed conformal symmetry
algebras generated by different pairs of the corresponding  ladder operators
and Hamiltonians of the rationally deformed 
conformal mechanics systems.
The commutators of the ladder operators 
$\mathcal{A}^\pm$, $\mathcal{B}^\pm$ and $\mathcal{C}^\pm$
with Hamiltonian operators are  given, respectively, 
by Eqs. (\ref{AaADarboux}), (\ref{BcalAladder}) and (\ref{C+-AAdefin})
with  $E^{\rm iso}=4$. The commutation relations   of the form
(\ref{AaADarboux}) also are valid for the case of the isospectral deformations
discussed in the previous section.
To write down the commutation relations between 
raising and lowering operators of the same type in general case, 
let us introduce the polynomial functions
\be
P_{n_+}(x)=\Pi_{k=1}^{n_+}(x-2n_k-1)\,,
\qquad R_{n_-}(x)=\Pi_{l=1}^{n_-}(x+2n_l+1),
\ee
where $n_k>0$ are the indices  of the corresponding seed states in the positive 
scheme and $-n_l<0$ are the indices 
of the seed states in the negative scheme. With this notation,
we have the relations $\A_{(+)}^+\A_{(+)}^-=P_{n_+}(L_0)$, 
$\A_{(+)}^-\A_{(+)}^+=P_{n_+}(L_{(+)})=P_{n_+}(L_{(-)}+2(n_-+n_+))$, 
and $\A_{(-)}^+\A_{(-)}^-=R_{n_-}(L_0)$,  $\A_{(-)}^-\A_{(-)}^+=R_{n_-}(L_{(-)})$. 
Then we obtain
\begin{equation}\label{AARH}
[\mathcal{A}^-,\mathcal{A}^+]=(x+1)(x+3) R_{n_-}(x)R_{n_-}(x+4) 
\big\vert_{x=L_{(-)}}^{L_{(-)}-4}\,,
\end{equation}   
\begin{equation}\label{BBPH}
[\mathcal{B}^-,\mathcal{B}^+]=(x+1)(x+3)P_{n_+}(x+4)P_{n_+}(x) 
\big\vert_{x=L_{(-)}+2N}^{x=L_{(-)}+2N-4}\,,
\end{equation} 
\begin{equation}\label{CCPRH}
[\mathcal{C}^-,\mathcal{C}^+]=R_{n_-}(x) P_{n_+}(x)\big\vert_{x=L_{(-)}}^{x=L_{(-)}+2N}\,,
\end{equation}
where $N=n_-+n_+$, and relation 
(\ref{AARH}) also is valid in the case of isospectral deformations.
In the case of the non-isospectral deformations given by the dual schemes
$(\ldots,2l_m-1,2l_m)\sim (\dots,-2l_m)$, the corresponding modified operators
(\ref{Cpmnew})  satisfy the commutation relation
\begin{equation}\label{tilCCPRH}
[\widetilde{\mathcal{C}}^-,
\widetilde{\mathcal{C}}^+]=(x+1)R_{n_-}(x) P_{n_+}(x+2)\big\vert_{x=L_{(-)}}^{x=L_{(-)}+2N-2}\,.
\end{equation}
Thus, 
in any rational deformation of the  conformal mechanics model we considered, 
each pair of the conjugate ladder operators of the types 
$\mathcal{A}^\pm$, $\mathcal{B}^\pm$ or $\mathcal{C}^\pm$ 
generates a non-linear deformation of the conformal $\mathfrak{sl}(2,\R)$ symmetry.
The commutation relations between ladder operators of different types of the form 
$[\mathcal{A}^\pm, \mathcal{C}^\pm]$, etc. not considered here are
model-dependent, and their taking into account gives rise naturally to different non-linearly extended  versions 
of the superconformal $\mathfrak{osp}(2|2)$ symmetry  \cite{InzPly+}.

 \section{Summary, discussion and   and outlook}\label{SectionSummary}

We studied rational deformations of the 
quantum conformal mechanics  model
and constructed  
complete sets  of  the spectrum-generating ladder operators for them
by applying the generalized DCKA 
 transformations to the quantum harmonic oscillator 
 and using the method of the dual ``positive" and ``negative" schemes 
and mirror diagrams.

The  ``positive" scheme is constructed by  selecting 
certain  $n_+$ physical states of harmonic oscillator as the seed 
states for  generalized DCKA transformation. 
In the ``negative" scheme we use some its $n_-$ non-physical states  
obtained from the corresponding eigenstates  with positive energies by the  
transformation $x\rightarrow ix$, $E\rightarrow -E$.
The two schemes are related 
by a kind of a ``charge conjugation"
presented by the mirror diagram.
They  generate the same deformed 
(or undeformed) conformal mechanics system 
modulo a nonzero 
 relative displacement 
of the spectrum
defined by a number of the seed states 
participating  in both dual schemes.
The set of the seed states in any of the  two dual schemes has to be chosen 
in such a way that the resulting system will be regular
on the positive half-line.
\vskip0.1cm

We showed that each \emph{isospectral} deformation of the isotonic oscillator
$L_m$ can be characterized by the ladder operators
$\mathcal{A}^\pm$
 being differential operators
of order $2(m+1)=2(n_-+1)$, which are the second order  ladder operators of $L_m$ 
Darboux-dressed by the conjugate pair of  the 
intertwining operators of order $m$  of the negative scheme.
These operators, like the second order ladder operators of $L_m$, 
form a spectrum-generating set of operators by means of which 
any state of the deformed system can be transformed into any other its state.
The picture, however, is completely different in the case of the gapped, \emph{non-isospectral},
deformations of  the  isotonic oscillators where it is similar 
to the picture  in rationally extended quantum harmonic oscillator systems 
\cite{CarPly2}.  The complete set of the spectrum generating 
operators is formed in this case 
by two pairs of the ladder operators $(\mathcal{A}^\pm,\mathcal{C}^\pm)$ 
or  $(\mathcal{B}^\pm,\mathcal{C}^\pm)$. 
Here the ladder operators $\mathcal{A}^\pm$ are constructed  
by Darboux-dressing of the second order ladder operators $(a^\pm)^2$
of the half-harmonic oscillator $L_0$ by the intertwining operators
of the negative scheme, while $\mathcal{B}^\pm$ are obtained
by Darboux-dressing of the same operators $(a^\pm)^2$
but by the intertwining operators of the positive scheme.
The  operators $\mathcal{A}^\pm$ and $\mathcal{B}^\pm$
 act similarly to the  usual ladder operators in the 
equidistant part of the spectrum but 
their action is different in a separated part.
The operators $\mathcal{A}^\pm$ detect all the 
separated states just by annihilating each of them.
The operators  $\mathcal{B}^\pm$
detect the number of low-lying separated 
valence bands in the spectrum where 
$\mathcal{B}^-$ acts as a lowering operator in each 
valence band  annihilating there the corresponding lowest state.
The raising operator $\mathcal{B}^+$ annihilates the highest 
state in each valence band. 
As a result,  in each valence band with  $l\geq 1$ states,
the action of the operators $\mathcal{B}^\pm$ is characterized
by the relations 
$(\mathcal{B}^\pm)^{l}=0$ typical for
spin-$s$ ladder operators  with $s=(l-1)/2$.
In dependence on the concrete non-isospectral  deformation,
sometimes the ladder operators $\widetilde{\mathcal{B}}^\pm$ 
of the lower differential order
can be constructed by the DC dressing 
of the ladder operators of some  
isotonic oscillator
$L_m$ with $m>0$, but their action 
on physical eigenstates of the deformed system 
is essentially the same as the action of  $\mathcal{B}^\pm$.
{}From the properties of  the ladder operators
$\mathcal{A}^\pm$ and $\mathcal{B}^\pm$ it is clear 
that they cannot connect the states from different 
valence bands, and the separated states cannot be connected by them with the states
from the equidistant part of the spectrum.
``Communication" between the indicated states
is provided by the ladder operators $\mathcal{C}^\pm$
constructed by gluing  the intertwining operators 
of the positive and negative schemes.
\vskip0.1cm
In any rational deformation of the  conformal mechanics model we considered, 
each pair of the conjugate ladder operators 
$\mathcal{A}^\pm$, $\mathcal{B}^\pm$ and  $\mathcal{C}^\pm$ 
(or  
$\widetilde{\mathcal{C}}^\pm$ in the case of the systems of the type 
$(\ldots,2l_m-1,2l_m)\sim (\dots,-2l_m)$)
generates some  non-linear deformation of the conformal $\mathfrak{sl}(2,\R)$ 
symmetry with a polynomial 
of order, respectively,  $2n_- -1$, $2n_+ -1$ and $n_- +n_+ -1$ 
(or $n_- +n_+ +1$) 
in the corresponding Hamiltonian of the system.
The commutation relations
 between  ladder operators  of different types are
model-dependent. 
The appropriate taking into account of them gives rise to different non-linearly extended  versions 
of the superconformal  $\mathfrak{osp}(2|2)$ symmetry \cite{InzPly+}.
\vskip0.1cm

It is  interesting  to note here that 
reflectionless (soliton and multi-soliton) quantum systems
can be  constructed by applying DC  transformations 
to the free quantum particle.
The latter system is peculiar in virtue of 
the presence in it of additional integral of motion 
being the momentum operator that distinguishes 
the left- and right-moving plane wave eigenstates 
corresponding to the doubly degenerate energy levels
of the free particle.
In our case  as analog of the free particle we have 
the  half-harmonic oscillator system $L_0$ 
characterized, as the harmonic oscillator,
by the equidistant spectrum and by 
the pair of the conjugate second order ladder operators 
associated with such a  peculiar nature of the spectrum.
Any reflectionless system
can be related  with the free particle 
by two different intertwining operators
whose action, however, does not produce any mutual
displacement. A similiar picture with two different
sets of the intertwining operators
acting without a relative
shift appears also 
in the isospectral pairs of periodic 
finite-gap systems. 
Because of the absence of the relative  
displacement in the action of different intertwining 
operators in reflectionless and finite-gap systems,
the analogs of our operators $\mathcal{C}^\pm$ built there 
by gluing different  intertwining operators are not
the ladder type operators.  Instead, they  are  
the Lax-Novikov integrals of motion of the corresponding 
quantum systems  related to the Korteweg-de Vries equation
and its hierarchy,   for the details see refs.
 \cite{exosusy1,exosusy2,exosusy3,exosusy4,AraPly}.
Moreover,  in this context we still have another interesting analogy. 
One can consider a particular class  of the
quantum reflectionless systems given by the
shape-invariant family of hyperbolic P\"oschl-Teller systems 
described by the potentials of the form $-m(m+1)/\cosh^2 x$
\cite{exosusy3}.
Their reflectionless properties together with the presence 
of a nontrivial Lax-Novikov integral in each  such a system 
are explained naturally in terms of their DC 
relationship with the free particle corresponding 
to the case $m=0$. If we change in the indicated potential 
the integer parameter $m$  for real parameter $\nu$, we loose the
connection with the free particle together with the reflectionless 
properties of the system. This analogy explains 
the peculiar nature of the rationally deformed 
conformal mechanics systems
characterized by integer parameter $m$ in comparison 
with the general (anyonic) case with $g=\nu(\nu+1)$,
$\nu>-1/2$, $\nu\notin \Z$.  
For the latter class of the systems 
it is impossible to assemble the ladder operators of the 
types  $\mathcal{A}^\pm$ and  $\mathcal{C}^\pm$, at least 
by the methods used here. 
Therefore, the problem of construction of the 
complete sets of the spectrum-generating ladder operators 
for such systems is open.
\vskip0.1cm

We  saw that the neighbour pairs of the isotonic oscillator
systems can be organized into the extended 
two-component  systems which are 
described by the $\mathcal{N}=2$ supersymmetry.
Such an extension can be realized in two different ways 
based on  physical (``positive") or non-physical (``negative")
eigenstates of the corresponding Hamiltonians of 
the subsystems. In correspondence with this, 
in the first case we have the unbroken $\mathcal{N}=2$ supersymmetry,
whereas in the second case the extended system 
is described by the broken supersymmetry. 
The interesting peculiarity we observed is that the products 
of the supercharges from the extended systems
with unbroken and broken supersymmetries 
produce the ladder operators of the conformal mechanics 
systems. 
A similar  observation was used recently in \cite{InzPly}
to explain the origin of the hidden superconformal symmetry 
of the quantum harmonic oscillator.

It  
would be interesting to investigate coherent states 
in the rationally extended quantum harmonic oscillator systems 
as well as in the rational deformations of the conformal mechanics  
considered here. 
The appearance of some peculiarities 
can be expected in such systems due
to the presence in them of different  sets of the 
ladder operators.

It also seems to be very interesting to identify 
some class of quantum physical systems  
in which the presence of more than one pair 
of the ladder operators would be essential.
One such a possibility could correspond 
to the problems related to quantum transitions.

\vskip0.2cm

\noindent {\large{\bf Acknowledgements} } 
\vskip0.2cm

LI acknowledges the CONICYT scholarship 21170053.
JFC and MSP acknowledge support from research projects
FONDECYT 1130017 (Chile),
Proyecto USA1555 Convenio Marco Universidades del Estado  (Chile), 
MTM2015-64166-C2-1 (MINECO, Madrid) and DGA E24/1
(DGA, Zaragoza).  JFC thanks
for the kind hospitality at Universidad de Santiago de Chile.
MSP is grateful for the warm hospitality at Zaragoza University.

 \appendix
\section{  Darboux transformations}\label{subSecDC}  

Here we briefly summarize  a general scheme of the Darboux transformations
and some their properties.
Let  $L_-=-\frac{d^2}{dx^2}+V_-(x)$ be a Schr\"odinger operator and 
 $\psi_*(x)$ be its eigenfunction,
  $L_-\psi_*=E_*\psi_*$.
We do not worry 
 about  the singular or non-singular 
nature of 
$L_-$ and the  physical or non-physical nature  of 
 $\psi_*(x)$, and
generate the first order differential operators
\be\label{defApsi}
A^-_{\psi_*}={\psi_*}\frac{d}{dx}\frac{1}{\psi_*}=\frac{d}{dx}+\mathcal{W}\,,\qquad
A^+_{\psi_*}=-\frac{1}{\psi_*}\frac{d}{dx}\psi_*=
-\frac{d}{dx}+\mathcal{W}\,,
\ee
where $\mathcal{W}=-\frac{\psi'_*}{\psi_*}$. 
The operator $A^+_{\psi_*}$ is
the formal adjoint of $A^-_{\psi_*}$,
and
$\ker\,{A^-_{\psi_*}}=\psi_*$, 
$\ker\,{A_{\psi_*}^{+}}=1/\psi_*$. 
We have a factorization relation
$A^+_{\psi_*}A^-_{\psi_*}=L_--E_*$,
and a representation of the potential $V(x)$ 
 in terms of the superpotential $\mathcal{W}$,
 $V_-=\mathcal{W}^2-\mathcal{W}'$.
 The alternate product 
$A^-_{\psi_*}A^+_{\psi_*}=L_+-E_*$
defines the associated
 Schr\"odinger operator $L_+=-\frac{d^2}{dx^2}+V_+$
 with the potential
 \be
 V_+=\mathcal{W}^2+\mathcal{W}'=V_-+2\mathcal{W}'=
 V_--2(\ln \psi_*)''\,,
 \ee
  and $L_+(\psi_*)^{-1}=E_*(\psi_*)^{-1}$.
 In dependence on the nature  of the wave function $\psi_*$,
 the so obtained operator $L_+$ can be singular or non-singular,
 and the wave function  $(\psi_*)^{-1}$ can describe
 a physical state or not.
 The operators $A^-_{\psi_*}$
 and $A^+_{\psi_*}$ intertwine the Schr\"odinger operators
 $L_-$ and $L_+$, 
 $A^-_{\psi_*}L_-=L_+A^-_{\psi_*}$, 
 $A^+_{\psi_*}L_+=L_-A^+_{\psi_*}$,
and provide us with the mapping between 
their physical and non-physical eigenfunctions.
Namely, if $E\neq E_*$ and $\psi_{-,E}$ is an eigenfunction of $L_-$,
$L_-\psi_{-,E}=E\psi_{-,E}$, then the function $A^-_{\psi_*}\psi_{-,E}$
is an eigenfunction of $L_+$ of the same eigenvalue $E$.
On the other hand, if $\psi_{+,E}$ is an eigenfunction of
$L_+$, $L_+\psi_{+,E}=E\psi_{+,E}$,  and  $E\neq E_*$,
then $A^+_{\psi_*}\psi_{+,E}$ is an eigenfunction  of $L_-$
of the same eigenvalue.
In the case of $E=E_*$ we have the relations 
$A^-_{\psi_*}\psi_{*}=0$,  $A^+_{\psi_*}\frac{1}{\psi_{*}}=0$,
and 
\be\label{Atildepsi}
A^-_{\psi_*}\widetilde{\psi_*}=(\psi_*)^{-1}\,,\qquad
A^+_{\psi_*}\widetilde{(\psi_*)^{-1}}=\psi_*\,,\qquad
\widetilde{\psi(x)}=\psi(x)\int^x\frac{d\xi}{(\psi(\xi))^2}\,.
\ee
Here 
$\widetilde{\psi(x)}$
is an eigenfunction  linearly independent 
from a solution $\psi(x)$ of the stationary Schr\"odinger equation
 $L\psi=E\psi$. 
So,  $\widetilde{\psi_*}$ in (\ref{Atildepsi})  
 is the eigenfunction of  eigenvalue $E_*$
 of $L_-$ to be linearly independent from $\psi_*$,
and  $\widetilde{(\psi_*)^{-1}}$
is the eigenfunction of the same eigenvalue $E_*$ of $L_+$
that is linearly independent from its eigenfunction $(\psi_*)^{-1}$.

One can apply iteratively various  Darboux transformations
to produce  a new system in $n$ steps. 
Such an  $n$-step process of construction of a final 
system can be presented equivalently as  a one-step
DC transformation \cite{MatSal}. 
The eigenstates $\psi_{[n],\lambda}$
of the final system are obtained by mapping 
the eigenstates $\psi_\lambda$  of the original system,
and can be presented in two equivalent forms\,:
\begin{equation}
\label{A8}
\psi_{[n],\lambda}=
\frac{W(\psi_1,\psi_2,\ldots,\psi_n,\psi_\lambda)}{W(\psi_1,\psi_2,\ldots,\psi_n)}
=\A_n^-\psi_\lambda\,.
\end{equation}  
A differential operator $\A_n^-$ of order $n$ is constructed iteratively,
\begin{equation}
\label{A9}
\A_n^-=A_n^-A_{n-1}^-\ldots A_1^-\,,\qquad A_i^-=
\A_{i-1}^-\psi_i\frac{d}{dx}\frac{1}{\A_{i-1}^-\psi_i}\,,
\end{equation}
where $i=1,\ldots,n$,  and  $A_0^-=1$. 
This operator intertwines the Hamiltonian of the initial system 
directly with the Hamiltonian of the final system.
Function (\ref{A8}) is a solution of the Schr\"odinger 
eigenvalue problem 
\begin{equation}
\label{A10}
\left(-\frac{d^2}{dx^2}+V_{[n]}(x)\right)\psi_{[n],\lambda}(x)=E_\lambda\psi_{[n],\lambda}(x),\,
\end{equation} 
where 
$V_{[n]}=V-2(\ln W(\psi_1,\ldots,\psi_n))''$
 is a potential of the 
corresponding final system.
The DC  transformation is defined by the choice
of the set of eigenfunctions $(\psi_1,\psi_2,\ldots,\psi_n)$  of 
the initial system
$L=-\frac{d^2}{dx^2}+V(x)$,
and we refer to the corresponding transformation as a 
 $(\psi_1,\psi_2,\ldots,\psi_n)$  scheme. 
 The change in the chosen order of the eigenfunctions $\psi_i$ 
 generates different intermediate operators  $\A^-_i$ and $A_i^-$  in factorization   
 (\ref{A9}),  but has  no effect on the  $n$-th order differential operator  $\A_n^-$
 and the potential $V_{[n]}$. 
 Nevertheless, we see that the order in which the seed states are  taken 
 is not unique, and  the intermediate schemes that occur in  the $n$-th order
 DC transformation  depend on the  order in  the 
 chosen set of eigefunctions  of the Hamiltonian operator of  initial system. 
The   kernel of $ \A_n^-$ 
is  spanned by the chosen set of the seed states,
$\ker \A_n^-=\text{span}\,\{\psi_1,\psi_2,\ldots,\psi_n\}$,
while $\ker \A_n^+=\text{span}\,\{\A_n^-\widetilde{\psi_1},\A_n^-\widetilde{\psi_2},\ldots,
\A_n^-\widetilde{\psi_n}\}$.
Eq. (\ref{A8}) can be presented in an equivalent form 
\begin{equation}
\label{schema2}
\psi_{[n],\lambda}=(A_n^-A_{n-1}^-\ldots A_2^-)(A_1^-\psi_\lambda)=
\frac{W(A_1^-\psi_2,\ldots,A_1^-\psi_n,
A_1^-\psi_\lambda)}{W(A_1^-\psi_2,\ldots,A_1^-\psi_n)}\,.
\end{equation}
Combining Eqs. (\ref{A8})  and (\ref{schema2}), we obtain 
 a new expression for Wronskian, 
\begin{equation}
\label{6.5}
W(\psi_1,\psi_2,\ldots,\psi_n,\psi_\lambda)=
{W(\psi_1,\psi_2,\ldots,\psi_n)}
\frac{W(A_1^-\psi_2,\ldots,A_1^-\psi_n,
A_1^-\psi_\lambda)}{W(A_1^-\psi_2,\ldots,A_1^-\psi_n)}\,.
\end{equation}

 Consider now the scheme $(\psi_1, \widetilde{\psi_1}, \psi_2,\ldots,\psi_n)$. 
We realize the first step on the basis of  the function $\psi_1$,  that produces
 the intertwining operator $A_1^-$. Its action  on $\widetilde{\psi_1}$
generates the function  $A_1^-\widetilde{\psi_1}=1/\psi_1$,
see Eq. (\ref{Atildepsi}).  According to (\ref{A9})
and (\ref{defApsi}), we obtain then
 $A_2^-=-(A_1^-)^{\dagger}$. When applied twice, 
 Eq. (\ref{schema2}) in this case 
 gives  us relations
\begin{equation}
\label{magicrule2}
\frac{W(\psi_1, \widetilde{\psi_1}, \psi_2,\ldots,\psi_n,\psi_\lambda)}{
W(\psi_1, \widetilde{\psi_1}, \psi_2,\ldots,\psi_n)}=
\frac{W(A_1^{\dagger}A_1^-\psi_2, \ldots, A_1^{\dagger}A_1^-\psi_n, 
A_1^{\dagger}A_1^-\psi_\lambda)}{W(A_1^{\dagger}A_1^-\psi_2, \ldots, A_1^{\dagger}
A_1^-\psi_n)}
=
\frac{W(\psi_2,\ldots,\psi_n,\psi_\lambda)}{W(\psi_2,\ldots,\psi_n)}\,.
\end{equation}
Here  we used the fact that the second order operator
$A_1^{\dagger}A_1^-$ is equal to the original Hamiltonian
shifted for some constant, 
and $\psi_i$, $i=2,\ldots,n$, are its eigenfunctions.
The second equality in (\ref{magicrule2}) is modulo 
some nonzero constant multiplier.
 Note that  for  a scheme $(\psi_1,\widetilde{\psi_1},\psi_2)$, by (\ref{A8}) 
  we obtain  the relation  $W(\psi_1,\widetilde{\psi_1},\psi_2)=W(\psi_1,
  \widetilde{\psi_1})A^{\dagger}_1A^-_1\psi_2$.  Since  $W(\psi_1,\widetilde{\psi_1}
  )=const\neq 0$,  the equality reduces to    $W(\psi_1,\widetilde{\psi_1},
  \psi_2)=C\psi_2$. This  corresponds to  a 
  particularly simple case 
  of a  relation $W(\psi_2,\ldots,\psi_n,\psi_\lambda)=
  W( \psi_1, \widetilde{\psi_1}, \psi_2,\ldots,\psi_n,\psi_\lambda)$
directly   following  from (\ref{magicrule2}).

\section{Equivalent  representations for eigenfunctions of $L_m$}\label{AppHerLag}
Eq.  (\ref{AmAmgen}) can be used  to establish the relationship
between  eigenfunctions (\ref{AAm}) 
of the isotonic oscillator given in terms of the Hermite polynomials
and their standard representation 
in terms of the generalized Laguerre polynomials.

The normalized 
form of the eigenstates (\ref{AAm}) 
is given by the relation $\widehat{\psi}_{m,l}=h_{m,l}\psi_{m,l}$
with normalization coefficients $h_{m,l}=2^{-(2m+l)}\pi^{-1/4}
\left((m+l)!(4m+4l+1)!/l!\right)^{-1/2}$.
On the other hand,
the $l$-th normalized eigenstate of the
isotonic oscillator (\ref{defisog}) with $g=m(m+1)$
is given by the relation \cite{Perel}
 $\widehat{\psi}_{m,l}=\left(2\cdot l!/\Gamma(m+l+\frac{3}{2})\right)^{1/2}(-1)^l
x^{m+1}\mathcal{L}^{m+1}_{l}(x^2)$,  and
we obtain the equality
\begin{equation}\label{AmH=L}
\A_{m}^-\left(H_{2n+1}(x)e^{-x^2/2}\right)=C_{nm}x^{m+1}\mathcal{L}^{(m+1/2)}_{n-m}(x^2)e^{-x^2/2},\qquad 
C_{nm}=(-1)^{n+m} 2^{2n+m+1}n!\,,
\end{equation}  
where $n=m+l \geq m$. Eq. (\ref{AmH=L}) can be extended for the
case $m=0$ if we put here $\A_0=1$ in correspondence with 
definition (\ref{A9}). 
 Multiplying this relation by $e^{x^2/2}$  from the left, we obtain
\begin{equation}\label{LnHn}
x^{m+1}\mathcal{L}^{(m+1/2)}_{n-m}(x^2)=C _{nm}^{-1}\mathcal{D}_m 
H_{2n+1}(x),\qquad \mathcal{D}_m=\left(\frac{d}{dx}-\frac{m}{x}\right)\ldots
\left(\frac{d}{dx}-\frac{1}{x}\right).
\end{equation}
Note that the  differential operator $\mathcal{D}_m$
of order $m$ appearing in (\ref{LnHn})  has exactly the form 
of the operator that intertwines the Hamiltonian 
operator of a free quantum particle on the positive half-line and 
that of Calogero model with coupling constant $m(m+1)$ \cite{MatPly}.
One can invert relation (\ref{LnHn}) by multiplying it from the left 
by the operator $e^{x^2/2}\A^+_m$
and applying Eq. (\ref{AmAmgen}). This gives the relation
\be\label{H2n+1Ln-m}
H_{2n+1}(x)=(-1)^{n+m}(n-m)! \, 2^{2n+m+1}
\mathcal{D}_m^\dagger\left(x^{m+1}
\mathcal{L}^{(m+1/2)}_{n-m}(x^2)\right).
\ee
Differentiating (\ref{H2n+1Ln-m}) 
and using the relation $(H_n(x))'=2n\,H_{n-1}(x)$
we also obtain the relation
\be\label{H2nLn-m}
H_{2n}(x)=(-1)^{n+m}(n-m)! \, 2^{2n+m}(2n+1)^{-1}\frac{d}{dx}
\mathcal{D}_m^\dagger\left(x^{m+1}
\mathcal{L}^{(m+1/2)}_{n-m}(x^2)\right).
\ee
Eqs.  (\ref{H2n+1Ln-m})  and  (\ref{H2nLn-m})
generalize  the well known relations 
of the case $m=0$ with $\mathcal{D}_0=\mathcal{D}_0^\dagger=1$
between  the generalized Laguerre and 
Hermite polynomials\,: 
$H_{2n+1}(x)=(-1)^{n} 2^{2n}n!\, 2x
\mathcal{L}^{(1/2)}_{n}(x^2)$ and 
$H_{2n}(x)=(-1)^{n} 2^{2n}n! 
\mathcal{L}^{(-1/2)}_{n}(x^2)$,
where for obtaining the last equality  we also
used the relation
$\frac{d}{dx}(x^\alpha \mathcal{L}^{(\alpha)}_n(x))=
(n+\alpha)x^{\alpha-1}\mathcal{L}^{(\alpha-1)}_n(x)$ with 
$\alpha=1/2$. 
\vskip0.1cm

\section{Proof of relation (\ref{reio1})}\label{AppB}

Our goal here  is to prove  relation (\ref{reio1}), for which  
the key point will be  the equality $H_{2n+1}(x)=x\mathcal{L}_{n}^{(1/2)}(x^2)$ 
modulo inessential  numerical constant. Consider the Wronskian identities 
$W(u(x)h_1(x),\ldots u(x)h_m(x))=u^m(x)W(h_1(x),\ldots,h_m(x))$, where $h_n(x)$,
 $n=1,\ldots, m$, and $u(x)$ are some arbitrary functions of $x$, and 
 $W(h_1(x^2),\ldots,h_{m}(x^2))=x^{m(m-1)/2}W(h_1(z),\ldots,h_{m}(z))|_{z=x^2}$.  
  Then for an arbitrary set of 
 odd states of the harmonic oscillator we have
\begin{equation}\label{B1}
W(\psi_{2n_1+1},\ldots,\psi_{2m_1+1})=x^{m(m+1)/2}
e^{-mx^2/2}g_{(-(2n_1+1),-(2n_2+1),\ldots,-(2n_m+1))}(x)\,,
\end{equation}     
where $g_{(-(2n_1+1),-(2n_2+1),\ldots,-(2n_m+1))}(x)=
W(\mathcal{L}_{n_1}^{(1/2)}(z),\ldots,\mathcal{L}_{n_m}^{(1/2)}(z))|_{z=x^2}$. 
Using the derivative relation for
 the generalized Laguerre polynomial, $\frac{d^k}{dz^k}(\mathcal{L}_{n}^{(\alpha)}(z))=
 (-1)^k \mathcal{L}_{n-k}^{(\alpha+k)}(z)$ if $k<n$, otherwise $\frac{d^k}{dz^k}
 (\mathcal{L}_{n}^{(\alpha)})(z)=0$, and also the relation $\mathcal{L}_{0}^{(\alpha)}=1$,
 for particular case when $n_j=j$, $j=0,1,\ldots m-1$,  one finds 
 that the  function $g_{(1,\ldots,2m-1)}$ is a determinant of some triangular matrix 
 with constant numbers in the diagonal, and  then  $W(1,\ldots,2m-1)=Cx^{m(m+1)/2}e^{-x^2/2}$. 
 Under the transformation $x\rightarrow ix$ the Wronskian (\ref{B1}) takes the form
\begin{equation}
\label{D13}
W(\psi_{2n_1+1}^{-},\psi_{2n_2+1}^{-},...,\psi_{2n_m+1}^{-})
= e^{mx^2/2}x^{m(m+1)/2}f_{(-(2n_1+1),-(2n_2+1),\ldots,-(2n_m+1))}(x)
\end{equation}
with $f_{(-(2n_1+1),\ldots,-(2n_m+1))}(x)=g_{((2n_1+1),\ldots,(2n_m+1))}(ix)$. 
This is just a particular example of a non-singular Wronskian of non-physical eigenstates. 
  
\section{List of some deformed conformal mechanics potentials}\label{list}
Here we display the explicit form of some deformations of the conformal mechanics 
potentials $V_m=x^2+\frac{m(m+1)}{x^2}$
discussed in the main part of the text\,:
\begin{eqnarray}
\label{V(-3,-7)}
&V_{(-3,-7)}=V_2-4+
  24 \frac{32 x^{10} + 240 x^8 + 528 x^6   - 840 x^4 - 3654 x^2  -2205  }{(8 x^6
  +60 x^4 + 126 x^2 +105 )^2}\,,&\\
\label{V(1,4,5)}
&V_{(1,4,5)}= V_1+ 6+ 
8 \frac{96 x^{10}-48 x^8-144 x^6-920 x^4+230 x^2-75}{(8 x^6-4 x^4+10 x^2+15)^2}\,,&\\
\label{V(1,5,6)}
&V_{(1,5,6)}=V_1+6 +16\frac{
256 x^{14}-960 x^{12}+1152 x^{10} -5712 x^8+432 x^6+9180 x^4-3960 x^2-675}{(16 x^8-48 x^6+72 x^4+60 x^2+45)^2}\,,&\\
\label{V(x)long}
&V_{(1,4,5,10,11)}=10+V_1+16 \,\frac{N(x)}{D(x)}\,, &
\end{eqnarray}
where 
\begin{eqnarray}
N(x)&=&-72937816875 + 359826563250 x^2 - 3559365463500 x^4 + 
     1124647108200 x^6 \nonumber\\
   && + 368202542400 x^8 + 1343539612800 x^{10} - 
     1951252934400 x^{12} + 822933619200 x^{14} \nonumber\\
   && - 1455591836160 x^{16} + 
     1118053063680 x^{18}  - 1055756298240 x^{20} \nonumber\\
   && + 535377653760 x^{22} 
     - 
     147987087360 x^{24} + 32552681472 x^{26} - 9212461056 x^{28}
     \nonumber\\
   && + 
     5052694528 x^{30}  - 1919090688 x^{32} 
     + 402784256 x^{34} - 
     49020928 x^{36} + 2621440 x^{38}\,,\nonumber\\
   D(x)&=&(467775 + 623700 x^2 - 
   374220 x^4 + 1995840 x^6 - 702240 x^8 + 94080 x^{10}\nonumber\\
   && + 146560 x^{12}  - 
   64512 x^{14} + 45824 x^{16} - 11264 x^{18} + 1024 x^{20})^2\,.\nonumber
\end{eqnarray}


\end{document}